# Interdisciplinary Papers Supported by Disciplinary Grants Garner Deep and Broad Scientific Impact

Minsu Park[1,2,3,4], Suman Kalyan Maity[2,3,4], Stefan Wuchty[5,6,7,8], Dashun Wang[2,3,4,9,*]

[1] Division of Social Science, New York University Abu Dhabi, Abu Dhabi, UAE

[2] Center for Science of Science and Innovation, Northwestern University, Evanston, IL, USA

[3] Northwestern Institute on Complex Systems, Northwestern University, Evanston, IL, USA

[4] Kellogg School of Management, Northwestern University, Evanston, IL, USA

[5] Department of Computer Science, University of Miami, Coral Gables, FL, USA

[6] Department of Biology, University of Miami, Coral Gables, FL, USA

[7] Sylvester Comprehensive Cancer Center, University of Miami, Miami, FL, USA

[8] Institute of Data Science and Computing, University of Miami, Miami, FL, USA

[9] McCormick School of Engineering, Northwestern University, Evanston, IL, USA

* Correspondence to: dashun.wang@northwestern.edu

**Short Title:** Grants and Interdisciplinary Advances

**Abstract:** Do interdisciplinary grants support high-impact interdisciplinary advances? We analyzed 350,000 grants from 164 agencies in 26 countries, along with 1.3 million resulting papers published between 1985 and 2009, to measure their interdisciplinarity and impact. Although interdisciplinary grants tend to produce interdisciplinary papers, which are generally associated with high impact, they yield fewer papers on average. Furthermore, the interdisciplinary papers they support tend to have substantially lower impact compared to those funded by disciplinary grants. In contrast, highly interdisciplinary papers supported by deeply disciplinary grants garner disproportionately more citations, both within their core disciplines and from broader fields. This impact advantage is not merely a consequence of funding size, reception of ideas within disciplinary boundaries, or collaborative formats. Amid rising support for interdisciplinary work, these results highlight the underexplored role of disciplinary grants in producing high-impact interdisciplinary advances, suggesting that interdisciplinary research may benefit from deep disciplinary expertise and investments.

**Teaser:** Disciplinary grants underpin impactful interdisciplinary research advances, highlighting the role of deep disciplinary expertise in science.

## Main Text

### Introduction

Many scientific challenges today, from climate change to global pandemics, require interdisciplinary approaches that integrate expertise and resources across diverse perspectives (*1*–*5*). Amidst the rapid growth in scale and complexity of the modern scientific enterprise (*3*, *4*, *6*), coupled with the increasing specialization of individual expertise (*7*, *8*), funding agencies and policymakers have been progressively focusing on grant programs that promote interdisciplinary work (*2*, *3*, *9*–*11*). Although funding plays a critical role in propelling scientific progress, our knowledge of how interdisciplinary grants shape the interdisciplinary research landscape remains limited. Yet, such understanding is essential for more productively supporting high-impact interdisciplinary endeavors, especially given the ensuing debates about the risks and benefits of interdisciplinary work among researchers and research institutions (*9*, *11*–*14*).

Prior studies have underscored the growing significance and impact of interdisciplinary work across scientific disciplines (*1*, *15*–*20*) by employing measures to quantify the interdisciplinarity of research *papers* (*19*–*22*). At the same time, another stream of research has examined the research outputs of *grants* (*23*–*26*). These studies typically rely on data from a single agency or country (*26*–*31*) and generally highlight the critical role of funding in propelling scientific progress, amidst the growing scale and complexity of science (*23*, *32*) and fiscal scarcity (*29*). While developing concomitantly, these two lines of research reveal an important gap in understanding the relationship between interdisciplinary grants and high-impact interdisciplinary advances they support.

This gap exists mainly due to the lack of a unified measurement approach to quantify the interdisciplinarity of *both* research grants and the resulting publications. To address this discrepancy, we combine data from two large-scale grant and publication databases—Dimensions (*33*) and the Microsoft Academic Graph (MAG) (*34*)—which are among the most comprehensive sources covering scientific grants and publications (*35*). We then introduce a new measurement framework and apply it to 350,000 grants from 164 funding agencies across 26 countries and 1.3 million papers that acknowledge these grants from 1985 to 2009 (see **Methods** for more details). This approach allows us to systematically examine the longitudinal changes in the interdisciplinarity of both research grants and papers across disciplines, as well as the relationships between grant interdisciplinarity and their supported publications, with a particular emphasis on the impact of these publications based on the interdisciplinary attributes of both the publications and their supporting grants.

The key technical challenge here is that while measuring the interdisciplinarity of papers is well established through bibliometric techniques based on references and citations (*15*, *16*, *19*–*22*, *36*), existing methods cannot directly be applied to grants, partly due to the lack of a consistent field classification scheme and standardized reference systems in grants. To tackle this challenge, we use field classifications of papers and their abstracts in the MAG dataset (**Fig. 1a**) to learn text representations of each scientific field (**Fig. 1b**) with a supervised topic modeling method, Labeled-Latent Dirichlet Allocation (Labeled-LDA; see **Methods**). Unlike methods that assign a single category, Labeled-LDA estimates word associations for each field, enabling us to calculate the probabilities of a grant's association across all potential fields based on its abstract (**Fig. 1c**). We validate our model through multiple approaches, including human ratings and out-of-sample predictions, demonstrating reliable model outputs (see Supplementary Note 3). Finally, to determine the probability that a given publication is associated with a particular field, we use the fraction of its references or citations in that field as a proxy of topical inspiration or appeal, respectively (**Fig. 1d**; see **Methods**), allowing us to express both grants and papers in probabilistic terms across multiple fields.

We then quantify the level of interdisciplinarity of individual publications and grants using the Rao-Stirling diversity as commonly operationalized in previous research (*15*, *16*, *19*–*22*, *36*). This measure incorporates three sets of information (**Fig. 1f**), including the number of research fields (*volume*; **Fig. 1c,d**), their relative distribution (*balance*; **Fig. 1c,d**), and their differences (*disparity*; **Fig. 1e**), on a scale from zero to one, where 0 indicates deeply disciplinary work and 1 indicates the highest level of interdisciplinarity (see **Methods** for more details). Together, these data and methods provide a unique opportunity to study grants and papers at a large scale under a unified field classification scheme.

## Results

**Figure 2a** shows an overall increasing trend in interdisciplinary research across the sciences over the past 25 years (see also Supplementary **Figs. S1**,**S2**), a result that is in line with previous observations (*1*, *11*, *15*). Notably, since the mid-1990s, papers that acknowledged grant support have exhibited a higher level of interdisciplinarity, hinting at the relevant role of funding in fostering interdisciplinary work (see Supplementary **Fig. S3** for the robustness of this result controlling for author prominence and team size).

We then examine 2,213,187 grant-paper pairs, capturing 1,293,934 publications and 350,526 supporting grants, and uncover two seemingly contradictory patterns. First, we observe that grants with higher interdisciplinarity tend to result in more interdisciplinary papers (**Fig. 2b** and Supplementary **Fig. S4**), and papers supported by interdisciplinary grants are found to attract citations from a wide range of disciplines (inset, **Fig. 2b**). Additionally, by calculating the paper-level hit rate, defined as the probability of a paper being in the top 5% of citations in its field and year (*37*), we find that highly interdisciplinary papers tend to be more impactful (**Fig. 2c** and Supplementary **Fig. S5**). These findings suggest that interdisciplinary grants appear to fulfill their intended goal of producing high-impact interdisciplinary advances (*9*, *17*, *18*).

However, when we consider all grants, regardless of whether they produced a paper, we find that interdisciplinary grants, on average, yield fewer papers compared to their disciplinary counterparts (**Fig. 2d**). Despite an overall impact advantage of interdisciplinary papers (**Fig. 2c**), publications supported by interdisciplinary grants tend to have a significantly reduced impact (**Fig. 2e**), surprisingly. We confirm the robustness of these results across different sample frames, including variations in funding agencies, time periods, disciplines, and countries (see Supplementary Note 5). These results paint a more nuanced picture of the role of interdisciplinary grants, suggesting that interdisciplinary papers supported by interdisciplinary grants appear to feature different characteristics than interdisciplinary papers in general. Together, the results in **Fig. 2** highlight the importance of considering the interdisciplinary orientation of *both* grants and their supported papers to understand the success of grants and their research outcomes, prompting us to further investigate the joint distribution of grant-paper pairs.

To that end, we categorize grant-paper pairs based on the interdisciplinary orientations of both papers and their supporting grants and report the average hit rate of papers in each category (**Fig. 3**). While the hit rate tends to increase with the interdisciplinarity of publications (**Fig. 2c** and Supplementary **Fig. S5**), **Fig. 3a** reveals that highest-impact papers are predominantly found in the upper left corner, suggesting that interdisciplinary papers supported by disciplinary grants tend to garner disproportionately high impacts. Note that disciplinary grants are less likely to produce interdisciplinary papers on average (**Fig. 2b** and Supplementary **Fig. S4**). Nevertheless, our findings indicate a systematic decline in the impact of papers as the interdisciplinarity of their supporting grants increases, even when controlling for the level of paper interdisciplinarity (**Fig. 3b** and Supplementary **Fig. S6**). We further split our samples by different funding agencies, time periods, disciplines, and countries and repeat our analyses, pointing to the same results (see Supplementary Note 5).

Overall, amidst the concomitant rise of both interdisciplinary research and funding, this result suggests that disciplinary grants appear to play an especially important role in producing high-impact interdisciplinary advances. At the same time, it also raises the question of why. One possibility is that disciplinary grants, born out of more established funding mechanisms (*2*, *10*), might receive larger funding support and therefore are more likely to produce higher-impact work. However, we find that interdisciplinary grants, on average, garner larger funding amounts compared to disciplinary grants (*38*) (**Fig. 4a**). Moreover, we observe increased publication productivity and impact for disciplinary grants even when controlling for funding size. Specifically, as the interdisciplinarity of grants increases, both the average number of outcome papers and their hit rate decrease sharply, regardless of grant size (**Fig. 4b,c** for large- and medium-sized grants, respectively). Note that this decreasing pattern is more pronounced with larger funding amounts while the baselines of productivity and impact rise with increasing funding size (see Supplementary **Fig. S7**).

Another potential explanation for the impact of disciplinary grants centers around the reception of ideas within disciplinary boundaries. For example, papers that were supported by deeply disciplinary grants may have home-field advantages, allowing them to acquire more citations, particularly from within their own fields. To investigate this point, we trace the top and bottom 25% of papers and supporting grants ranked by their interdisciplinarity. Then, we calculate the average number of citations that these papers received from within and outside their own field. **Fig. 4d** reveals that papers supported by disciplinary grants (top and bottom left) indeed enjoy a home-field advantage, as they accumulate more citations than expected from their own field. More importantly, interdisciplinary publications supported by disciplinary grants (top left) tend to garner higher impact not just within their core disciplines but also from broad and distant fields. This finding suggests that interdisciplinary papers supported by disciplinary grants are associated with both deep and broad scientific impacts.

Finally, as teams are increasingly responsible for producing high-impact advances (*7*, *37*, *39–43*), we examine the organizations of collaborative grants and ask whether specific combinations of collaborative formats are particularly suited for the production of high-impact interdisciplinary publications. For instance, highly disciplinary grants from distant disciplines may foster interdisciplinary advances by combining deep disciplinary expertise across disparate scientific fields. In other words, individual grants may be deeply disciplinary, but they may be combined with those from other disciplines to enable interdisciplinary efforts. To investigate this, we consider papers that acknowledged support from multiple grants. For each paper, we compute both the average interdisciplinarity of the supporting grants and the average disciplinary distance between them. We then categorize these papers into four groups based on the interdisciplinarity and distance scores of their supporting grants. These groups represent different collaborative grant formats: proximate disciplinary grants; distant disciplinary grants; proximate interdisciplinary grants; and distant interdisciplinary grants (from left to right in **Fig. 4e**). Upon comparing the impact of papers supported by these four distinct collaborative formats, we find that papers garner the highest impact when they are highly interdisciplinary and supported by multiple disciplinary grants that are proximate in their intellectual space. Conversely, the impact of papers decreases when supported by distant disciplinary grants, and it sinks for publications resulting from collaborations involving distant interdisciplinary grants. These patterns are robust after controlling for a range of funding- and author-level factors (see Supplementary Note 4). Overall, our results suggest that while distant disciplinary grants can span broader intellectual terrains, closely-related disciplinary grants tend to be more effective in producing impactful interdisciplinary work, further highlighting the significant role of disciplinary grants in fostering high-impact interdisciplinary advances (see Supplementary Note 5 for the robustness of our key results across different funding agencies, time periods, disciplines, and countries).

## Discussion

Despite these findings, several limitations suggest avenues for further study. First, this paper focuses on grants' outcomes in terms of papers and citations. While these are major outputs, funders often also emphasize broader impacts, such as outreach, practical applications, and policy relevance, which are not captured by our publication-based measures (*11*). Future work may also integrate diverse forms of interdisciplinary support, including seed grants, training programs, and targeted faculty hiring, to encompass a wider range of outcomes. Second, our data trace grant outcomes through grant acknowledgments in the paper. While this is a common practice in similar studies, some grants may be acknowledged tangentially or inconsistently. One open question is whether one can refine acknowledgment analyses by distinguishing relative contributions or validating acknowledgments through complementary data. Finally, our analysis focuses on empirical relationships between interdisciplinary grants and the papers they support. Future work may attempt to clarify the mechanisms beneath these empirical regularities, which may arise from multidimensional forces, including social, institutional, and cultural contexts that shape knowledge production (*44*, *45*). Taken together, these considerations underscore that while our findings highlight the significant role of disciplinary grants in fostering interdisciplinary research, it should not be viewed as a dismissal of interdisciplinary grants, which remain essential for nurturing diverse research programs and cross-disciplinary collaborations. Rather, our findings serve as a starting point, prompting richer, multi-dimensional evaluations of interdisciplinary initiatives and their broader impacts.

Overall, our results show that the broad and deep impacts of disciplinary grants are not simply a consequence of funding size, reception of ideas within disciplinary boundaries, or collaborative grant formats. Even with comparable funding resources, disciplinary grants tend to be more effective in producing high-impact interdisciplinary advances than their interdisciplinary counterparts and seem especially powerful when paired with other closely related disciplinary grants. A contributing factor may be the tendency of interdisciplinary work, when fueled by disciplinary grants, to draw attention and garner citations from both its core field and broad external fields. While our analyses are correlational by nature and do not allow causal interpretations, these results align with the view that "narrow work has broad impact" (*15*) and further emphasize the advantage of deep disciplinary expertise in the ambit of research (*7*, *46*). At the same time, amidst the broad shifts toward interdisciplinary sciences (*1*, *15*, *16*), our findings highlight the enduring challenges of interdisciplinary work, suggesting that the fruits of interdisciplinary programs are not always guaranteed. While interdisciplinary grants appear to produce intended outcomes, i.e., papers with high interdisciplinarity, we find that highly interdisciplinary grants tend to yield fewer total papers and a reduced probability of producing highly impactful papers, despite having larger funding on average.

While unveiling the often-overlooked role of disciplinary grants in producing key interdisciplinary insights, our findings further reflect the substantial costs and risks of interdisciplinary research, highlighting the need to manage tensions among different disciplinary and professional approaches (for research communities) and integrate deep disciplinary expertise in driving interdisciplinary work (for individual researchers and teams). Challenges may arise from the difficulties in collaborative relationships (*47*–*49*), developing a common language (*48*, *50*, *51*), focusing on a shared perspective from disparate viewpoints, cultures, and traditions (*2*, *50*, *52*, *53*), and evaluating interdisciplinary work (*54*). The power of disciplinary grants in producing interdisciplinary advances that garner deep and broad impacts, therefore, raises important questions for academics, funders, and policymakers on how to best unleash the full potential of interdisciplinary research and programs.

## Materials and Methods

**Dataset of research grants and articles.** We draw upon the Dimensions dataset (*33*), which tracks scientific publications and the grants that they acknowledge. Our analysis focuses on grants that were awarded after 1985, capturing 350,526 grants and 1,293,934 resulting papers that were published before 2009 (to allow time for citations to accumulate, given that our citation data cutoff is in 2020). To compute interdisciplinarity measures, we only include papers with at least one reference and one citation. Overall, these papers and grants cover 292 fields and 164 funding agencies across 26 countries. We further complement this dataset with abstracts, fields of study labels, and reference and citation information from corresponding papers by merging the Dimensions data with the Microsoft Academic Graph (MAG) dataset (*34*). Note that we provide further details on the extensive coverage and comprehensiveness of our data sources, which surpass those of other widely-used databases (*35*), along with discussions addressing potential concerns in Supplementary Note 1.

**Fields of study.** In defining research fields, we align with the notion of topical coherence as the systematic production of knowledge, particularly as manifested in content (*55*–*57*). Similarly, contrary to views that define interdisciplinarity by the disciplinary backgrounds of grant recipients or paper authors, we focus on the thematic content of proposals and publications. This perspective is crucial for understanding the thematic continuity between a grant's objectives and the resulting research output, highlighting the tangible link between the nature of a grant and the characteristics of the research it supports.

Among the widely-used large-scale databases, including Scopus and the Web of Science (WoS), MAG uniquely classifies papers into fields of study based on their content, irrespective of the source of publication. Also, given the limitations of journal-based categorizations, especially in fields where journals are not the primary medium of scientific communication (e.g., Computer and Information Sciences), MAG's content-based classification offers a more comprehensive scheme with comparable coherence. This approach, grounded in the indexing and classification of publications in MAG, offers a common basis for our analysis.

In this study, we use the field information from the MAG dataset, which assigns each paper to at least one research field using a four-level hierarchical classification. Specifically, we associate each publication with 292 level-1 fields, which are comparable to the granularity of classifications in other popular bibliographic databases such as WoS. The validity of our approach is demonstrated by the similarity of our results on the longitudinal trends of interdisciplinarity of publications (**Fig. 2a**) with those reported in Gates et al. (2019) (*15*), which relied on WoS data.

**Field representation in a grant (Labeled-LDA).** A key empirical challenge in quantifying the interdisciplinarity of grants is systematically assigning grants to the research fields they belong to. Here, we use a new measurement approach using Labeled-LDA (*58*), allowing us to estimate the probability that a given grant is associated with a particular scientific field based on its abstract. Specifically, we train our model on a sample of 572,302 paper abstracts and their one or more field-of-study labels. We obtain this sample through random sampling of 1 million papers from the MAG dataset but exclude papers without level-1 field label or with abstracts under 100 words. The resulting model constructs a one-to-one correspondence between latent topics and labels, enabling us to learn a probability distribution of word-field associations. We validate our model through manual inspections of these word-field associations as well as out-of-sample classification tasks (see Supplementary Note 3). Additionally, by applying our methods to papers, we find that the distances between fields computed by the Labeled-LDA method and citation patterns (described in the section on 'Distance between fields' below) exhibit a moderate positive correlation (Pearson's $r = 0.451$, $P < 0.001$), showing general consistency between our method and the literature. These validation results also indicate that the used field categories are conceptually coherent and align well with general understandings of fields of study, thereby supporting the validity of MAG's field categories.

In applying the trained Labeled-LDA model to individual grant abstracts, we calculate the probability of a grant being associated with specific scientific fields. In our assessment of grant interdisciplinarity, we re-normalize the field probabilities by excluding those with the lowest probability score, deemed irrelevant, to vary the number of pertinent fields (capturing the notion of volume and variety in the defined interdisciplinarity below). The estimation of field probabilities of grants is analogous to the vector of probabilities that a publication is associated with research fields as described below (see the section on 'Field representation in a paper'). Note that we replicate the main results without the re-normalization process. Furthermore, our approach is not confined to a mere classification task. It is highly adept at estimating document-label *relevance* in probabilities across multiple pre-defined labels, which is particularly useful when a coherent labeling scheme is absent in one system (i.e., 'grants') but can be extrapolated from another (i.e., 'papers'). This capability enables us to analyze both research grants and publications under a unified field classification scheme (see Supplementary Note 3 for more details).

**Field representation in a paper.** Following previous research (*15*, *16*), we use a paper's references to estimate interdisciplinary inspiration and its citations to estimate the interdisciplinary impact of a paper. We first represent each publication by a vector over 292 scientific fields, $p$. By considering all references of a paper, we compute the paper's probability to belong to field $i$ ($p_i$) as a fraction of references that are associated with field $i$. We apply the same process when we consider citations of a paper.

**Distance between fields.** As scientific fields vary in their proximity, we compute the distance between fields by estimating the overall knowledge stock within a discipline. In particular, we consider the cumulative reference or citation vectors $v_i$ over a set of $n$ papers within the field $i$, where $v_i = \{p_{1,i}, \ \ldots, p_{n,i}\}$. The distance, $d_{ij}$, is then defined as the cosine distance between fields $i$ ($v_i$) and $j$ ($v_j$), $d_{ij} = 1 - \frac{v_i \cdot v_j}{|v_i| \cdot |v_j|}$. Here, fields whose papers have very similar reference or citation patterns have a small distance $d_{ij} \approx 0$, while fields whose papers have very different reference or citation patterns have a large distance $d_{ij} \approx 1$. Using a $M \times N$ discipline proportion matrix of $p_i$ values (for each row, i.e., paper, $\sum_i p_i = 1$), we compute the cosine distances between all field pairs. Note that the distances between fields that were determined from references and citations are highly similar (Pearson's $r = 0.978$, $P < 0.001$), indicating the robustness of this approach.

**Grant and paper interdisciplinarity (Rao-Stirling diversity).** Our definition of interdisciplinarity emphasizes 'diversity' and 'coherence,' reflecting the integration of knowledge from multiple research fields and the intensity of relations between these knowledge bodies (*57*). Numerous metrics, including network and entropy measures, have been proposed to assess interdisciplinarity, possibly yielding inconsistent results (*57*, *59*–*61*). However, consensus among scholars stipulates that simply counting the number of disciplines that occur in references and citations is inadequate for properly quantifying interdisciplinarity. A more comprehensive approach considers not only the count but also the relative proportion of each discipline (capturing entropy) and the distance between disciplines (reflecting the intrinsic dissimilarity between disciplines) (*15*, *19*, *61*, *62*). For example, a paper primarily referencing computer science and information science is less diverse than one that equally draws from both computer science and economics. Consequently, the Rao-Stirling Diversity has emerged as a common measure to quantify interdisciplinary research (*15*, *16*, *19*, *22*, *36*, *61*). The Rao-Stirling index of a grant or a paper is defined as $RS(p) = 2 \cdot \sum_{i \neq j} p_i p_j d_{ij}$, where $p_i$ ($p_j$) is the probability that the underlying grant (or paper) is associated with discipline $i$ ($j$) while $d_{ij}$ is the distance between discipline $i$ and $j$. An RS score of 0 reflects a lack of interdisciplinarity (i.e., all references, citations, or grants are from the same discipline), whereas an RS score of 1 corresponds to the highest level of interdisciplinarity.

To provide more comprehensive understanding, in Supplementary Note 2, we discuss discrepancies in various measurement approaches of interdisciplinarity and potential confounding factors related to our interdisciplinary measure.

**Acknowledgements:** This work uses data sourced from Dimensions.ai through researcher access plans. The major part of this research is carried out during M.P.'s research fellowship at the Center for Science of Science and Innovation at Northwestern University.

**Funding:** D.W. is supported by the National Science Foundation grant TF-2404035. S.W. is supported by National Science Foundation grants 2123635 and 2210023. M.P. was partially supported by the NYUAD Center for Interacting Urban Networks (CITIES), funded by Tamkeen under the NYUAD Research Institute Award CG001.

**Author contributions:** D.W., S.W., and M.P. conceived the project and designed the experiments; D.W. secured data; M.P. performed empirical analyses with the help from D.W., S.W., and S.K.M; M.P., S.W., and D.W. discussed and interpreted results; M.P., S.W., and D.W. wrote and edited the manuscript.

**Competing interests:** The authors declare no competing interests.

**Data and materials availability:** The data and code necessary to reproduce the main and supplementary results will be shared in a permanent repository.

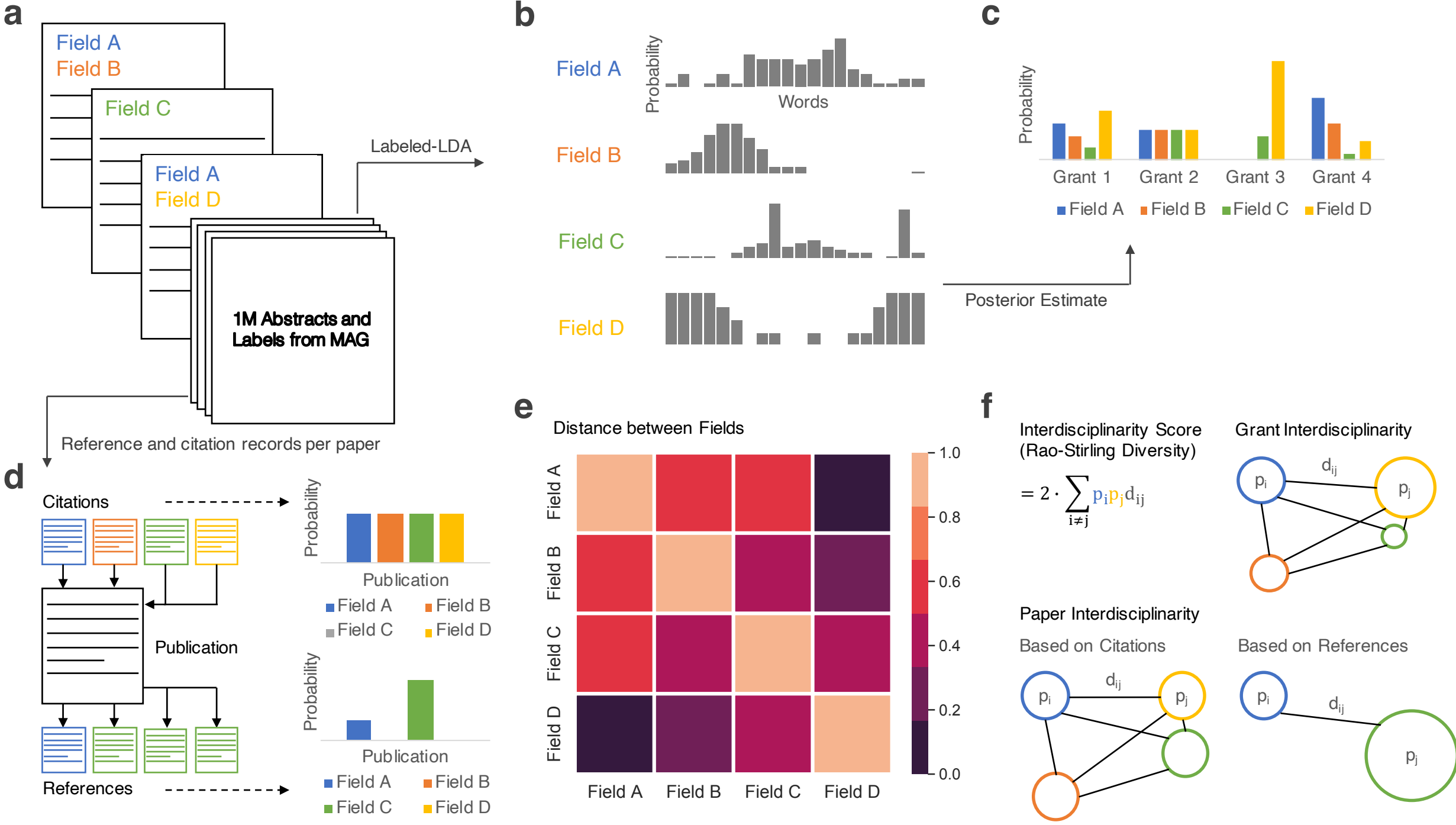

**Fig. 1 | Quantifying the level of interdisciplinarity of individual publications and grants.** Major publication databases assign each paper to certain scientific fields, while grant classifications are specific to individual funding agencies. **a,** We collect abstract and field labels of each publication from the Microsoft Academic Graph (MAG) database to build a semi-supervised topic model. **b,** Based on a large-scale representative sample of publications, we associate each word in an abstract with the field of study labels of the corresponding paper and *vice versa* using Labeled-Latent Dirichlet Allocation (Labeled-LDA), allowing us to obtain a robust representation of word associations for each scientific field. **c,** Using our trained Labeled-LDA model, we estimate the posterior probabilities that a grant belongs to a given scientific field based on the word distribution in the corresponding grant abstract. **d,** In turn, we calculate the probabilities that a paper belongs to a scientific discipline based on the fields of referenced and citing publications, respectively. **e,** We estimate the distances between scientific fields using cosine similarity between the reference (or citation) vectors that we obtain from corresponding publications in each field. Note that the reference- and citation-based distances are highly correlated with each other (Pearson's $r = 0.978$, $P < 0.001$), suggesting that our result is insensitive to the measurement specification. **f,** Based on the field-relevance probabilities of grants and papers computed in **c,d** and distances between fields computed in **e**, we calculate the level of interdisciplinarity of each grant and paper with the Rao–Stirling diversity measure.

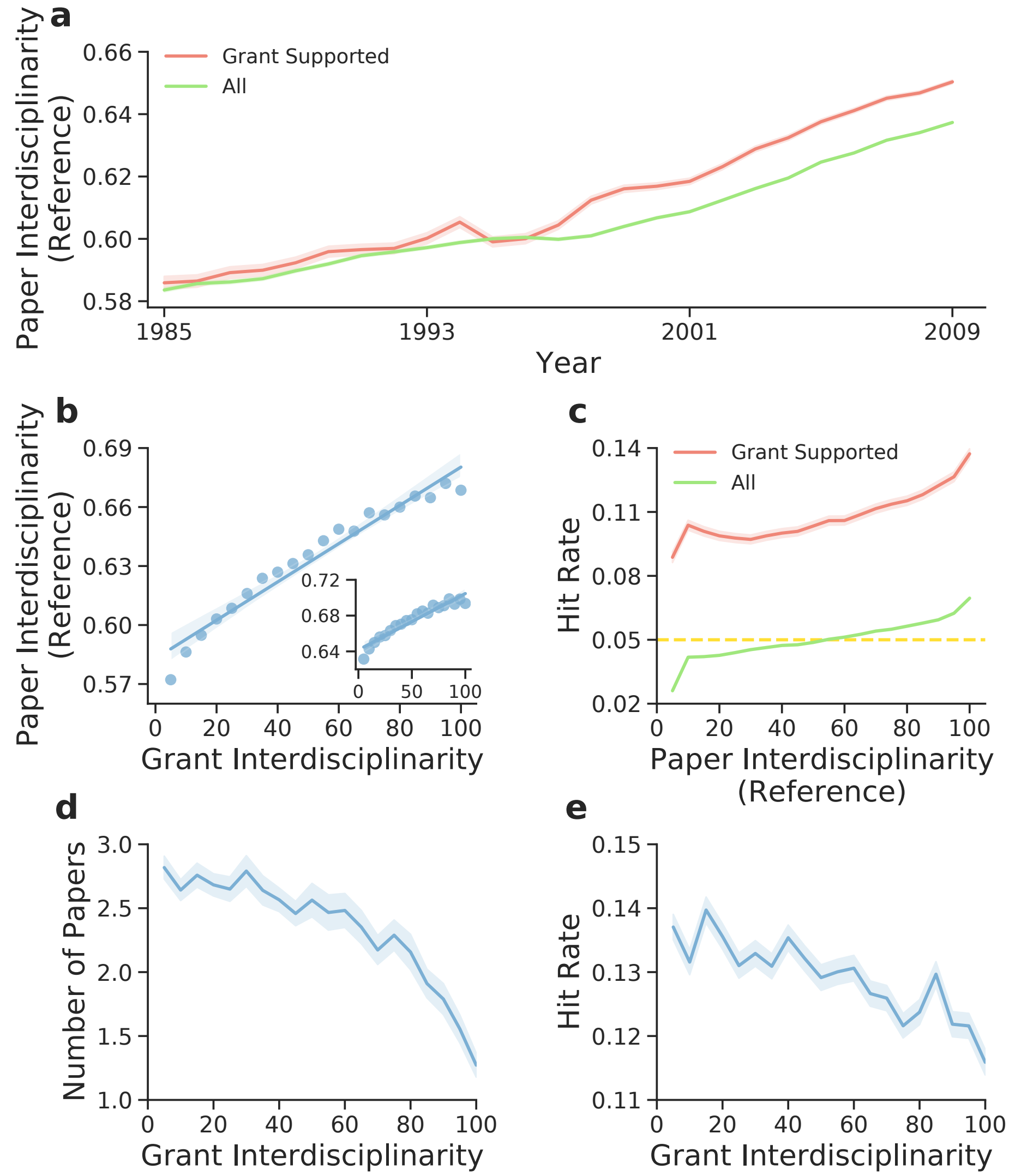

**Fig. 2 | Impacts of interdisciplinary grants. a,** Paper interdisciplinarity has been rising steadily from 1985-2009, and the increase of interdisciplinarity is more pronounced when we consider papers with grant support. **b,** Paper interdisciplinarity, as measured through paper references, increases as a function of the interdisciplinarity of supporting grants. Inset shows similar results when we consider paper interdisciplinarity based on citations. **c,** Papers with high interdisciplinary inspirations (i.e., reference-based paper interdisciplinarity) have a higher chance to be hit papers (dashed line as the baseline). This relationship also holds for grant-supported papers. The number of papers resulting from a grant (**d**) and the propensity to produce hit papers (**e**) systematically decrease as grant interdisciplinarity increases.

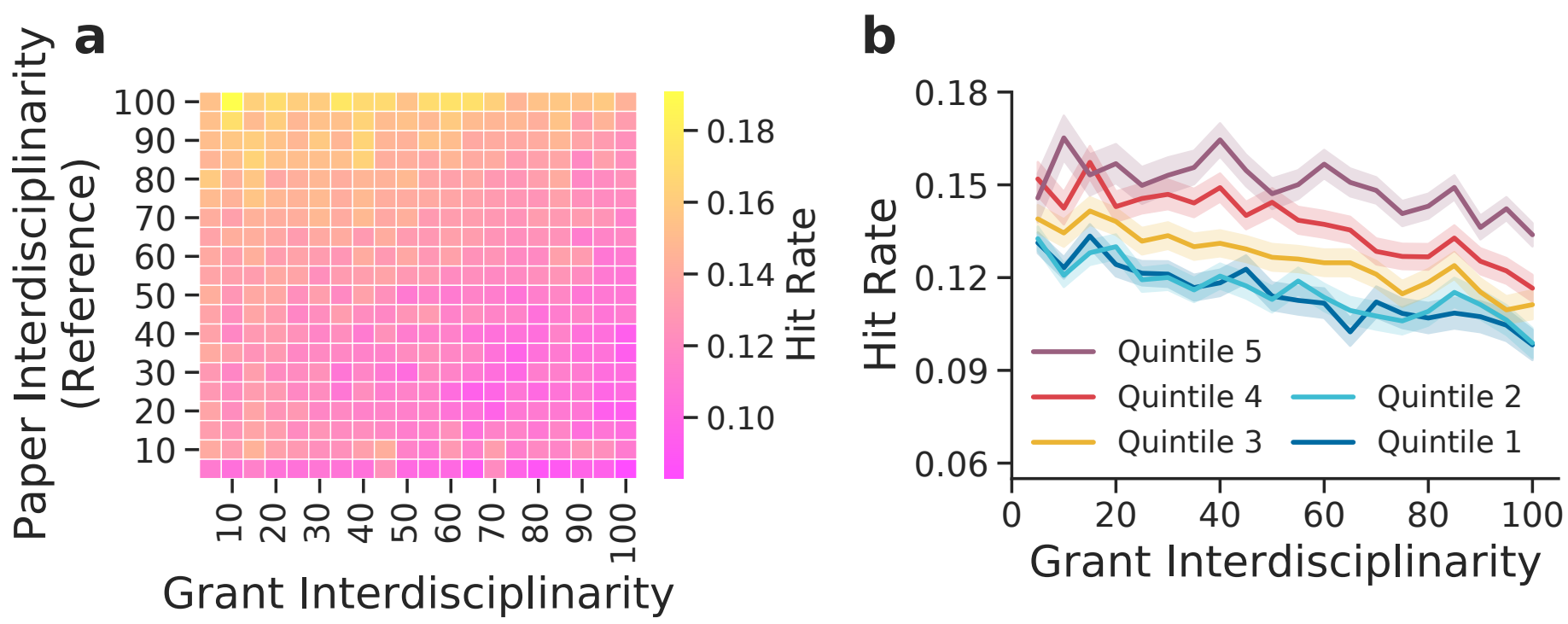

**Fig. 3 | Impact of interdisciplinary papers as a function of grant interdisciplinarity. a,** Interdisciplinary papers from more disciplinary grants tend to be associated with higher impact. **b,** While the baseline average of impacts increases with paper's interdisciplinarity (from Quintile 1 to Quintile 5), interdisciplinarity grants have an overall reduced probability of supporting impactful papers when controlling for papers with the same level of interdisciplinarity (based on references).

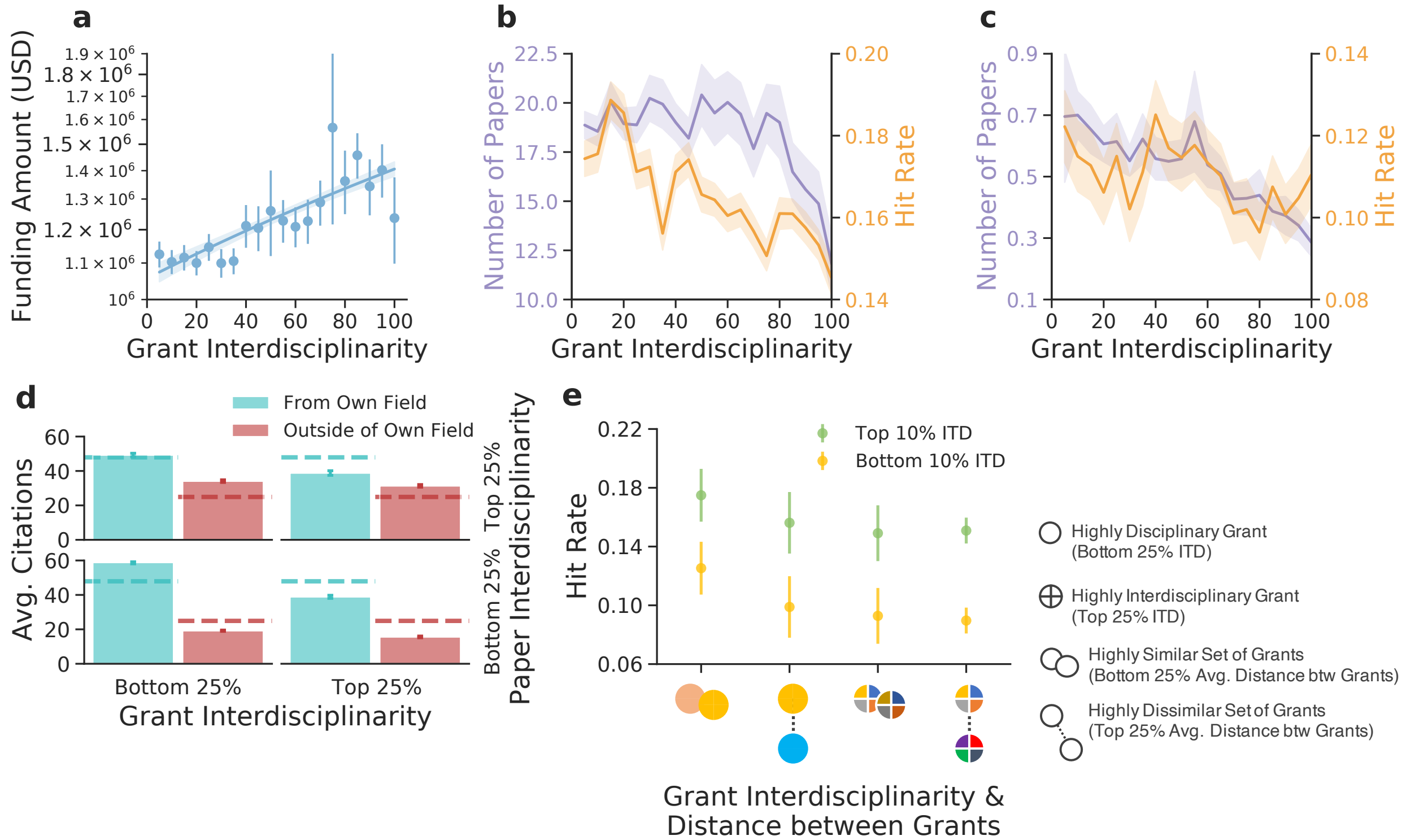

**Fig. 4 | Disciplinary grants and high-impact interdisciplinary papers. a,** Interdisciplinary grants tend to feature larger funding amounts. **b,** Considering only grants with large funding amounts (top 10%), we observe a sharp decline in both productivity (purple) and impact (orange) as a function of grant interdisciplinarity. **c** shows a similar pattern of diminishing returns when we focus on grants with median funding amounts (middle 10%). **d,** Interdisciplinary papers supported by disciplinary grants (top left) tend to have a similar or higher number of citations than baselines (dashed lines) both from inside and outside of their own fields. Other papers attract more citations than the random baseline either from their own field (disciplinary papers supported by disciplinary grants; bottom left), outside their own field (interdisciplinary papers supported by interdisciplinary grants; top right), or neither (disciplinary papers supported by interdisciplinary grants; bottom right). In **e,** we consider sets of the top and bottom 10% interdisciplinary papers based on their references that were supported by multiple grants. We calculate the distance between grants and further divide the groups of publications into sets of highly (dis)similar pairs of (inter)disciplinary grants. We find that high-impact interdisciplinary papers tend to acknowledge the support of closely related disciplinary grants.

Supplementary Information for

# Interdisciplinary Papers Supported by Disciplinary Grants Garner Deep and Broad Impact

Minsu Park[1,2,3,4], Suman Kalyan Maity[2,3,4], Stefan Wuchty[5,6,7,8], Dashun Wang[2,3,4,9]

[1] Division of Social Science, New York University Abu Dhabi, Abu Dhabi, UAE

[2] Center for Science of Science and Innovation, Northwestern University, Evanston, IL, USA

[3] Northwestern Institute on Complex Systems, Northwestern University, Evanston, IL, USA

[4] Kellogg School of Management, Northwestern University, Evanston, IL, USA

[5] Department of Computer Science, University of Miami, Coral Gables, FL, USA

[6] Department of Biology, University of Miami, Coral Gables, FL, USA

[7] Sylvester Comprehensive Cancer Center, University of Miami, Miami, FL, USA

[8] Institute of Data Science and Computing, University of Miami, Miami, FL, USA

[9] McCormick School of Engineering, Northwestern University, Evanston, IL, USA

**This document includes:**

## Supplementary Note 1: Data Sources and Potential Biases

### 1.1 Potential Biases in the Coverage of Dimensions and Microsoft Academic Graph

Our study uses two primary data sources, Dimensions (1) and the Microsoft Academic Graph (MAG) (2). Dimensions is currently the only data source that offers extensive information on grants across various agencies, making it an invaluable resource for our analysis. A recent paper (3) highlighted that Dimensions encompasses 83.7% of grant-paper pairs from the NSF and 99.7% from the NIH, underscoring the dataset's coverage. On the other hand, MAG stands out for its comprehensive coverage of the scientific literature. In direct comparisons, such as the study by Visser et al. (2021) (4), MAG has been shown to surpass other widely-used databases like Scopus, Web of Science (WoS), and Crossref, particularly in terms of comprehensiveness within specific publication types and the breadth of covered publication types. Notably, MAG includes not only conventional journals but also conference proceedings—a primary publication avenue in fields like Computer and Information Sciences—and non-English publications.

In our dataset, out of 5,796,221 papers supported by grants from 1985 to 2009 in Dimensions, only 360,985 papers could not be matched with their corresponding publications in MAG, resulting in a 93.8% match rate. We are, therefore, confident that the linkage between the two data sources does not introduce discernible biases. Furthermore, our various robustness checks, described in the Supplementary Note 4, involve applying specific filters to focus on select segments of the grants and publications. The consistency in our results across these subsamples bolsters the reliability of our conclusions and addresses concerns about potential biases stemming from the different coverages of the data sources.

### 1.2 Completeness of MAG and Its Implications

While no dataset is perfectly exhaustive, MAG stands out for its extensive coverage and comprehensiveness of scientific literature, as mentioned above (4). The completeness of MAG may vary over time—with more recent years being more complete—and across different countries of publication. However, our analyses have shown consistent results across various time periods and countries, as reported in Supplementary Note 4. This robustness alleviates concerns about the potential impact of any incompleteness in the dataset on the results and conclusions of our study. Furthermore, using the Web of Science (WoS) dataset, a recent study by Gates et al. (2019) (5) identified longitudinal trends in interdisciplinarity that closely mirror our findings. This congruence indicates that the primary observations and conclusions of our study are likely to be replicated with other extensive datasets.

## 1.3 Availability of Dimensions Data and Reproducibility

Regarding reproducibility concerns, it is important to note that MAG is publicly available, facilitating the replication of our study. The Dimensions dataset, while accessible, is not freely available. Researchers or institutions with access to Dimensions can directly reproduce our results. For those without access, SciSciNet (3) offers a viable alternative. It provides access to data from the National Science Foundation (NSF) and the National Institutes of Health (NIH), enabling researchers to validate key components of our findings. Although SciSciNet may not cover the entire scope of our analysis, it is particularly useful for replicating aspects of our study that rely primarily on NSF and NIH grants.

## 1.4 Geographical Distribution of the Grants and Papers

A significant portion of our sample is concentrated in a number of countries, with the United States (62.9%), Japan (11.2%), China (8.8%), United Kingdom (3.9%), Belgium (3.9%), Russia (2.8%), and the Czech Republic (1.5%) being the most represented. Each of these countries accounts for more than 5,000 grants paired with at least one paper outcome. We acknowledge that this distribution may suggest a bias towards the scientific outputs and funding patterns prevalent in developed countries, particularly those with substantial investments in scientific research. However, this concentration is reflective of the global research landscape, where a few countries dominate scientific production and funding. Thus, we believe that the trends and patterns identified in our study offer valuable insights that are broadly applicable, especially in understanding the dynamics of interdisciplinary research and its impact. The cross-cultural robustness of our findings, as demonstrated in Supplementary **Fig. S11**, suggests that the principles and relationships we have explored are likely be the case in other national contexts, though the specific magnitudes may differ.

To further bolster the cross-cultural validity of our results, we have extended our analysis to additional countries, including Japan, the United Kingdom, and Belgium, each with over 10,000 grant-paper pairings in our dataset, allowing for comprehensive replication. It is important to note a few exceptions that do not detract from our broader conclusions: In the UK, contrary to the general trend, we observe a marginal decline rather than an increase in average funding amounts as grant interdisciplinarity rises. Similarly, in Belgium, the anticipated decrease in publication numbers with increasing grant interdisciplinarity is not observed; instead, these numbers appear to have plateaued. This consistency in findings across different countries further clarifies the generalizability of our results, particularly emphasizing their relevance to developed countries with significant scientific investments.

### 1.5 Distribution of Grants and Papers by Funding Types

The distribution of grants and papers by funding types, particularly concerning the career stages of the investigators, can be another concern. For example, grants awarded to early-career researchers versus established scientists may exhibit significant differences in interdisciplinarity. Since our dataset does not provide detailed information on whether specific types of funding were designated for researchers at varying career stages, this limitation precludes us from directly analyzing the impact of career stage on research interdisciplinarity.

Nevertheless, we have indirectly examined this aspect through the lens of funding size as a proxy for the type and target of the funding. As reported in our study, while funding size does have a positive and significant impact on the success of publications, our analysis reveals that the influence of the interdisciplinarity of grants and papers on a paper's success is largely independent of the funding size (see Supplementary Note 4). This finding suggests that the interdisciplinarity inherent in the research, both from the perspective of grants and papers, plays a significant role in determining research success, along with other important factors like funding amount. Despite the absence of specific data on funding types for different career stages, this aspect of our analysis provides valuable insights into the role of interdisciplinarity in research success.

## Supplementary Note 2: Potential Issues of the Interdisciplinarity Measurement

### 2.1 Discrepancies in Measurement Approaches

In the field of interdisciplinarity research, discrepancies in results across different measures often arise from the chosen unit of analysis. For example, assessing interdisciplinarity at higher groupings, such as fields or journals, can introduce complexities due to different methods of measuring aggregate interdisciplinarity. Our study, however, focuses on individual papers and grants as the primary units of analysis. In this specific context, different measures of interdisciplinarity are expected to yield similar outcomes (6), thereby ensuring a reasonable assessment of interdisciplinarity and avoiding the potential pitfalls of aggregate measurements.

Additionally, it is also important to note that the Rao-Stirling index, our chosen measure, is sensitive to the choice of the distance parameter. To mitigate this issue, we have used cosine distance, a recommended approach that suppresses the sensitivity to distance parameters (7, 8).

## 2.2 Relationships between Interdisciplinarity and Potential Confounding Factors

### 2.2.1 Number of References and Citations

Given that the reference and citation counts of papers have also been increasing over time (9, 10), somewhat mirroring the longitudinal trends in paper interdisciplinarity we observe (**Fig. 2a** and Supplementary **Fig. S1**), there may be concerns that our reported longitudinal trends are predominantly driven by the number of references cited in a paper and the number of citations it receives. However, Gates et al. (2019) (5) demonstrated that the Rao-Stirling Index, when conditioned on the number of references or citations, exhibits consistent trends over time, albeit with varying baselines. This means that the average trends over all articles (i.e., what we present) align with the qualitative trends observed when considering the number of references or citations. To maintain clarity in our presentation, we have chosen to report the average statistic for all articles, without differentiating based on reference or citation count.

To further ensure that the relationship between the impact and interdisciplinarity of papers is not confounded by the number of references, we added the number of references as a control variable in the regression and confirm that the results are identical (see Supplementary Note 4).

### 2.2.2 Number of Authors

The trend of increasing authorship in papers over time (11, 12) can raise questions about the stability of interdisciplinarity measures for analysis on longitudinal trends. To address this, we incorporated various team-related factors into our regression analysis, including proxies for team size such as the number of authors, grants, and institutes (see Supplementary Note 4 for more details). We also considered factors influencing success, like cross-cultural collaboration, as indicated by the number of funding countries. Our analysis shows that the effect sizes of our primary variables—paper interdisciplinarity, average grant interdisciplinarity, and grant-grant similarity—remain robust, suggesting that the increasing number of authors does not unduly influence the stability of interdisciplinarity patterns in our study.

### 2.2.3 Number of Research Fields

As the number of fields associated with a paper has been increasing over time—roughly about 3% from 1985 to 2009 (see Supplementary **Fig. S8a**)—and the average increase of the number of fields associated with a paper is also highly correlated with the average increase of the interdisciplinarity over time ($r = 0.929$, $P < 0.001$), there may be concerns about potential biases in our interdisciplinarity index, possibly explaining the increasing trend of interdisciplinarity shown in **Fig. 2a**. However, the interdisciplinarity conditioned on the number of fields associated with a paper shows the same trends over time (only the baseline is different; see Supplementary **Fig. S8b**). In other words, the average trends over all articles (i.e., what we present) matches the qualitative trends conditioned on the number of fields associated with a paper. This means that the interpretation of results and conclusion drawn from those results are not affected by this fact. Therefore, to simplify our presentation, we only report the average statistic over all articles regardless of the number of fields associated with a paper.

## Supplementary Note 3: Validation of Labeled-LDA Model

To estimate grant-field associations, we trained a Labeled-LDA model (13) using 573,302 abstracts randomly selected from the MAG database (2), each associated with one or more of 292 field labels. Our choice of Labeled-LDA over other methods was informed by several considerations: (i) a widely-used text classifier such as Support Vector Machine (SVM) (14) does not provide a word-to-category probability distribution, which is essential information for interpretability of the classification outcomes; (ii) such classifiers typically assign each document to a single label, which is not optimal for multi-label classifications; and (iii) while standard topic models like Latent Dirichlet Allocation (LDA) (15) compute a word-to-category distribution, they are not designed for such computation with pre-defined classes, since they are unsupervised learning methods.

While approaches based on a Large Language Model (LLM) can offer an advanced classifier, they require the use of an arbitrary threshold to filter out irrelevant categories and do not inherently provide a probability distribution. In contrast, by employing a Dirichlet prior, topic models inherently produce a probability distribution across topics. Additionally, given that each document is represented by a limited number of topics (i.e., field labels), the less relevant topics are consequently assigned minimal or zero probability scores. This model behavior allows for the intuitive identification of irrelevant fields without the need for an explicit threshold, while offering an accurate and probabilistic depiction of relevance across various fields (13). This model behavior is particularly crucial for accurately and probabilistically depicting relevance across various fields.

Since we aimed to build a model that learns word distributions with document-specific label distributions (unlike SVM), incorporates supervision by constraining the topic model to use only those topics that correspond to a document's observed label set (unlike LDA), and automatically select relevant topics without introducing an arbitrary threshold (unlike LLM-based approaches), Labeled-LDA was an appropriate option. The resulting model constructed a one-to-one correspondence between latent topics and labels (fields), from which a word-label (i.e., word-field) distribution could be learned (see Supplementary **Table S1** for the top 10 words in each field by probability and FREX score (16), a harmonic mean of relative frequency and exclusivity).

We evaluated the quality of our Labeled-LDA model using both human-centered and automated approaches as described below:

- **Direct human ratings:** We first estimated topic quality through direct ratings. We required an adequate number of raters to ensure sufficient statistical power (at least $1 - \beta = 0.9$) and draw meaningful conclusions from human annotations. Following Hoyle et al. (2021) (17), we had eighteen independent raters for each topic to obtain significance at $\alpha = 0.05$. These raters, all of whom held graduate degrees and resided in the United States, were recruited via Amazon Mechanical Turk and were compensated at a rate of 1.5 USD per survey, equivalent to roughly 18 USD/hour.

  For the evaluation, we randomly selected 20 fields. For each field, we pulled the top 10 words by probability and the top 10 words by FREX score, resulting in a total of 20 words. Each rater was provided with ten field-word sets (as shown in Supplementary **Table S1**), randomly chosen from the 20 fields, and was asked to give the topic quality on a conventional three-point ordinal scale ranging from 1 (not very related) to 3 (very relevant) (17, 18). The average quality score was $2.414 \pm 0.245$, with all scores falling within the range of 1.722 to 2.778, indicating the high reliability of our topic model.

- **Out-of-sample prediction.** We further examined the model's text-field representation by testing multi-label classification performance on out-of-sample paper abstracts and their corresponding field labels. We randomly sampled 5,000 papers as testing data, yielding an average precision of 0.461, significantly higher than the random baseline of 0.006.

- **Distance between predicted and ground truth labels.** The above evaluation is highly conservative because we consider only the field with the highest probability as the predicted label. In practice, a paper can have multiple labels (from 1 to 4), and an answer is counted as correct only when all labels match perfectly. Moreover, our Labeled-LDA model infers field-to-grant associations using 292 field labels, rather than mapping a text to a single field. Consequently, even if the top-ranked predicted field does not match the ground truth, lower-ranked fields may still capture relevant conceptual similarities. For instance, if the predicted field is 'Humanities' while the ground truth is 'Classics,' the inference may still be valid because these fields overlap conceptually.

  To address this nuance, we also examined whether our model reasonably captures the perceived similarity or distance between fields by computing the distance between predicted and ground truth labels, leveraging field-field similarities inferred by Labeled-LDA. Specifically, we determined pairwise topic similarity based on word probability distributions, where a distance of 0 represents a perfect match and 1 indicates that the predicted field is conceptually the most distant field to the ground truth. The mean distance was 0.311, significantly smaller than the random baseline of 0.606 ($P < 0.001$), indicating that our model's field-of-study estimation aligns substantially well with actual field labels, thus supporting its validity.

In sum, these validations ensure that the topic representation reasonably captures the nuances of each field and infers the field-grant associations effectively.

## Supplementary Note 4: Multivariate Analysis

We further examined the relationship between a paper's impact and the interdisciplinarity of both the paper and the supporting grants using Ordinary Least Squares (OLS) regression. We used 10-year citations (C10) to gauge a paper's impact, and OLS served as our primary estimation method. Supplementary **Table S2** summarizes the results.

In the interdisciplinarity model (Model 1), we included variables for a paper's interdisciplinary inspirations, the mean interdisciplinarity of the supporting grants, and the average field similarity among those grants. The non-interdisciplinarity model (Model 2) included grant-related variables—number of supporting grants, number of institutes, number of funding countries, and total funding amounts (USD)—to account for structural factors influencing citation counts, along with the number of authors to capture author-related conditions. We excluded additional grant- and author-related variables (e.g., number of grant investigators and institutes involved in the paper) to avoid multicollinearity. The combined model (Model 3) integrated both sets of variables to assess their net effects. Note that we emphasize Model 1 as the primary explanatory model for the intellectual and conceptual impacts of interdisciplinarity, whereas Model 3 serves as a robustness check that incorporates structural factors.

To address right-skewed distributions of citation counts, as well as grant- and author-related attributes, we log-transformed C10, total funding amounts, and the number of grants, institutes, and authors. We also included year and discipline fixed effects as dummy variables to control for time trends and discipline-specific factors tied to interdisciplinarity. Finally, we standardized all continuous independent variables to help comparison across variables and mitigate potential multicollinearity. We evaluated multicollinearity using the Variance Inflation Factor (VIF) and no severe multicollinearity was detected (VIF values ranged from 1.04 to 1.37, below the common threshold of 5).

As shown in Supplementary **Table S2**, Model 1 indicates that a paper's level of interdisciplinary inspiration has a positive effect on its success ($P < 0.001$). Meanwhile, the mean interdisciplinarity of supporting grants shows a negative association ($P < 0.001$), suggesting that papers backed by more disciplinary research programs tend to be more successful. Additionally, the average distance among supporting grants is negatively associated with paper success ($P < 0.05$), implying that papers benefit more from closely related grants than from highly diverse ones—results consistent with our main findings.

In the non-interdisciplinarity model (Model 2), the number of authors and the number of funding countries both exhibit significantly positive coefficients ($P < 0.001$), aligning with previous work (19). However, the coefficient direction of the number of institutes deviates from earlier findings. This model also supports our finding that the number of grants and total funding amount are positively associated with citation counts ($P < 0.001$).

When both sets of variables are combined in Model 3, most coefficients remain stable, but the number of institutes becomes insignificant. Meanwhile, the consistent coefficients and significance of interdisciplinary variables—and the explanatory power of Model 1 ($R^2 = 0.097$)—indicate that interdisciplinarity effects persist even alongside structural predictors. The modest increase in $R^2$ from Model 2 (0.135) to Model 3 (0.139) reflects shared variance among predictors rather than a lack of explanatory power. This outcome is unsurprising, given that interdisciplinary grants generally secure larger funding (**Fig. 4a**) and previous research shows that larger grants and teams can boost paper impact (12, 19). Together, these findings underscore the complex interplay between structural factors, interdisciplinarity, and citation impact, warranting deeper investigation of their interrelationships.

To further validate our results, we estimated an additional set of models using Negative Binomial regression, which is more appropriate for count data with overdispersion ($M = 54.97$, $SD = 110.42$). Although the log-transformation of C10 in OLS effectively corrected skewness, Negative Binomial regression allows us to directly model citation counts while accounting for variance inflation. The only difference from the OLS specification is that C10 remains in its raw count form rather than being log-transformed. The results, presented in Supplementary **Table S3**, remain highly consistent with the OLS findings. The consistency across both modeling approaches—OLS with log-transformed C10 and Negative Binomial regression with raw citation counts—demonstrates that interdisciplinarity effects on citation impact are not artifacts of model selection but hold across different specifications.

## Supplementary Note 5: Robustness of the Results

To confirm the robustness of our results based on all of the grant and publication data from 1985 to 2009, we replicated our key analyses in **Figs. 2b,d**, **Fig. 3a**, and **Figs. 4a,d**, considering different funding agencies (NIH and NSF; Supplementary **Fig. S9**), time windows (before and after 2000; Supplementary **Fig. S10**), countries (United States and China, among others; Supplementary **Fig. S11**), and disciplines (Applied Sciences, Formal Sciences, Humanities, Social Sciences, and Natural Sciences; Supplementary **Fig. S12-14**). While these results are almost identical to the observations we reported in the main text based on the entire data, there are a few key exceptions in cross-discipline analysis, summarized as follows:

- In Formal Sciences (including Computer Science and Mathematics), we found that more interdisciplinary grants tend to produce a higher number of papers.

- In Humanities (encompassing Art, History, and Philosophy), we observed that highly disciplinary papers supported by highly disciplinary grants tend to receive more citations, both from their core field and externally. Additionally, the number of papers plateaus even as the interdisciplinarity of grants increases. However, due to limited data points in this discipline, these statistics should be interpreted with caution.
- For other disciplines, such as Applied Sciences (Business, Engineering, Materials Science, and Medicine), Social Sciences (Economics, Geography, Geology, Sociology, Political Science, and Psychology), and Natural Sciences (Biology, Chemistry, Environmental Science, and Physics), all the patterns align closely with our main findings.

## Supplementary References

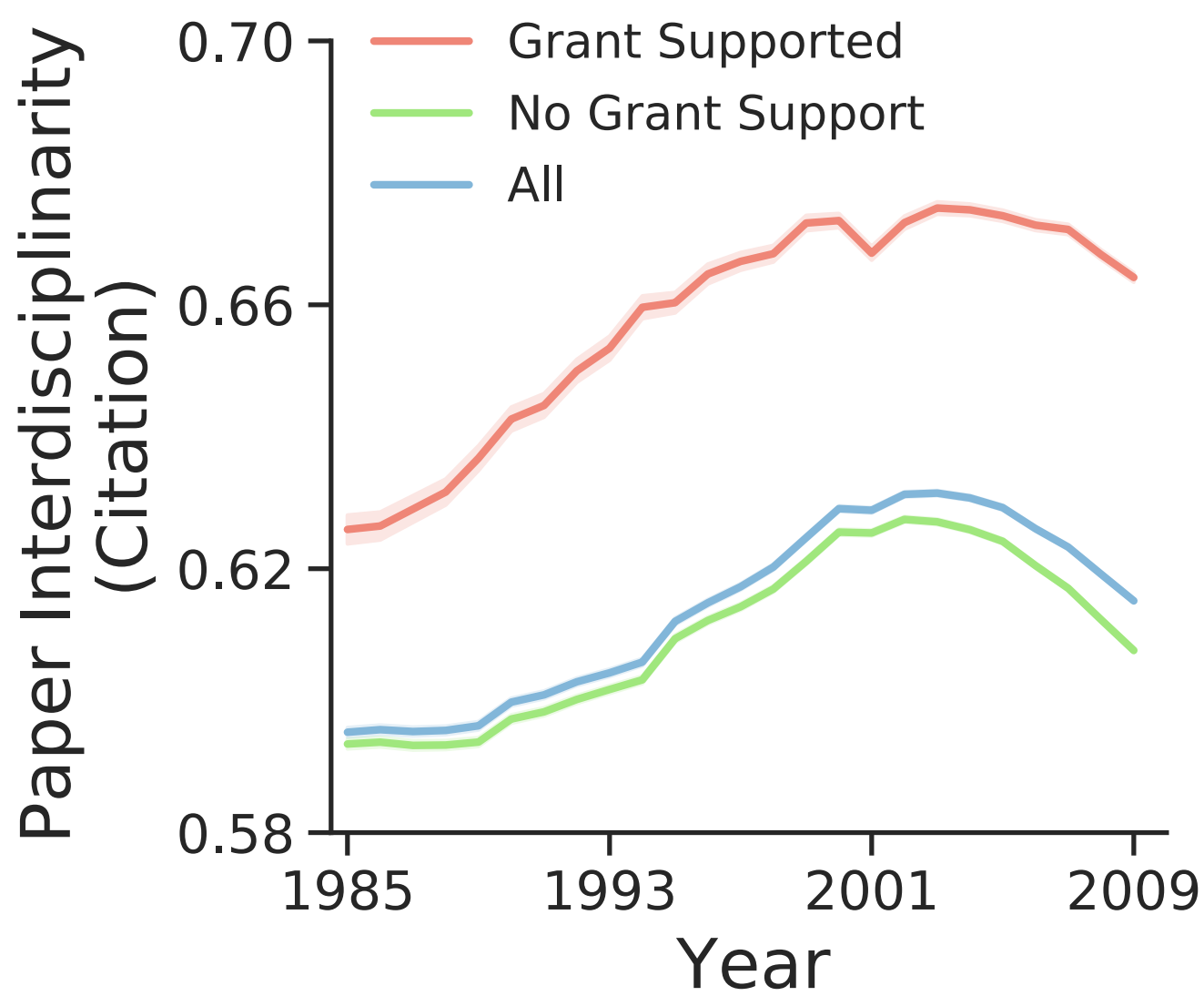

**Supplementary Fig. S1 | Mean interdisciplinarity of papers based on citations increased over time.** We obtained similar results to those in **Fig. 2a** when we considered the interdisciplinarity of papers through their received citations as a proxy for a publication's broad appeal. In turn, we observed a recent decline in the interdisciplinarity of citations that papers garnered that appeared to be roughly independent of grant support. Such an observation may be rooted in the fact that cross-discipline citations tend to emphasize older papers, suggesting that papers need time to accumulate more interdisciplinary citations(20).

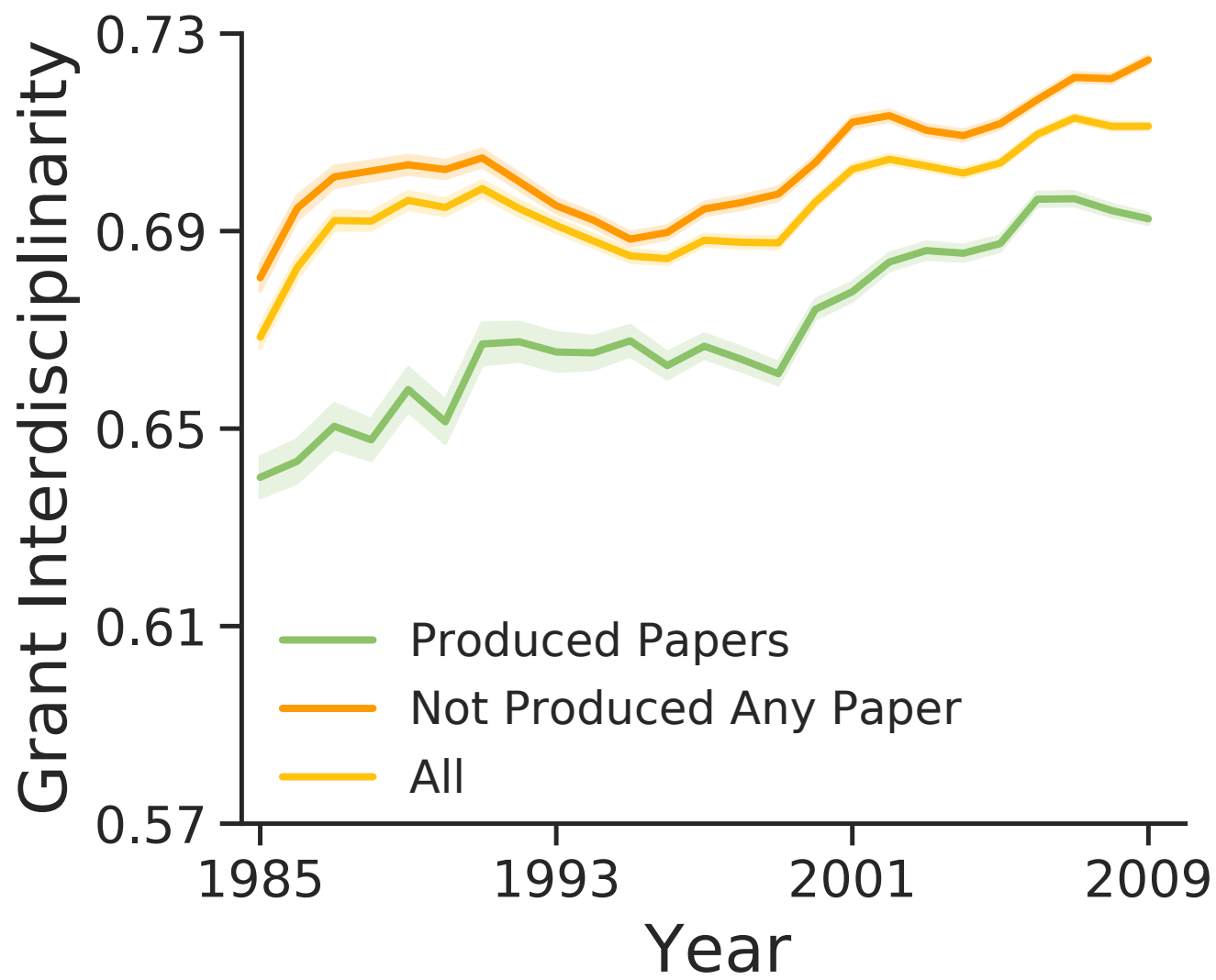

**Supplementary Fig. S2 | Mean interdisciplinarity of grants increased over time.** While more interdisciplinary grants have been awarded over time, we curiously find that grants that produced published papers are less interdisciplinary than grants that produced no papers.

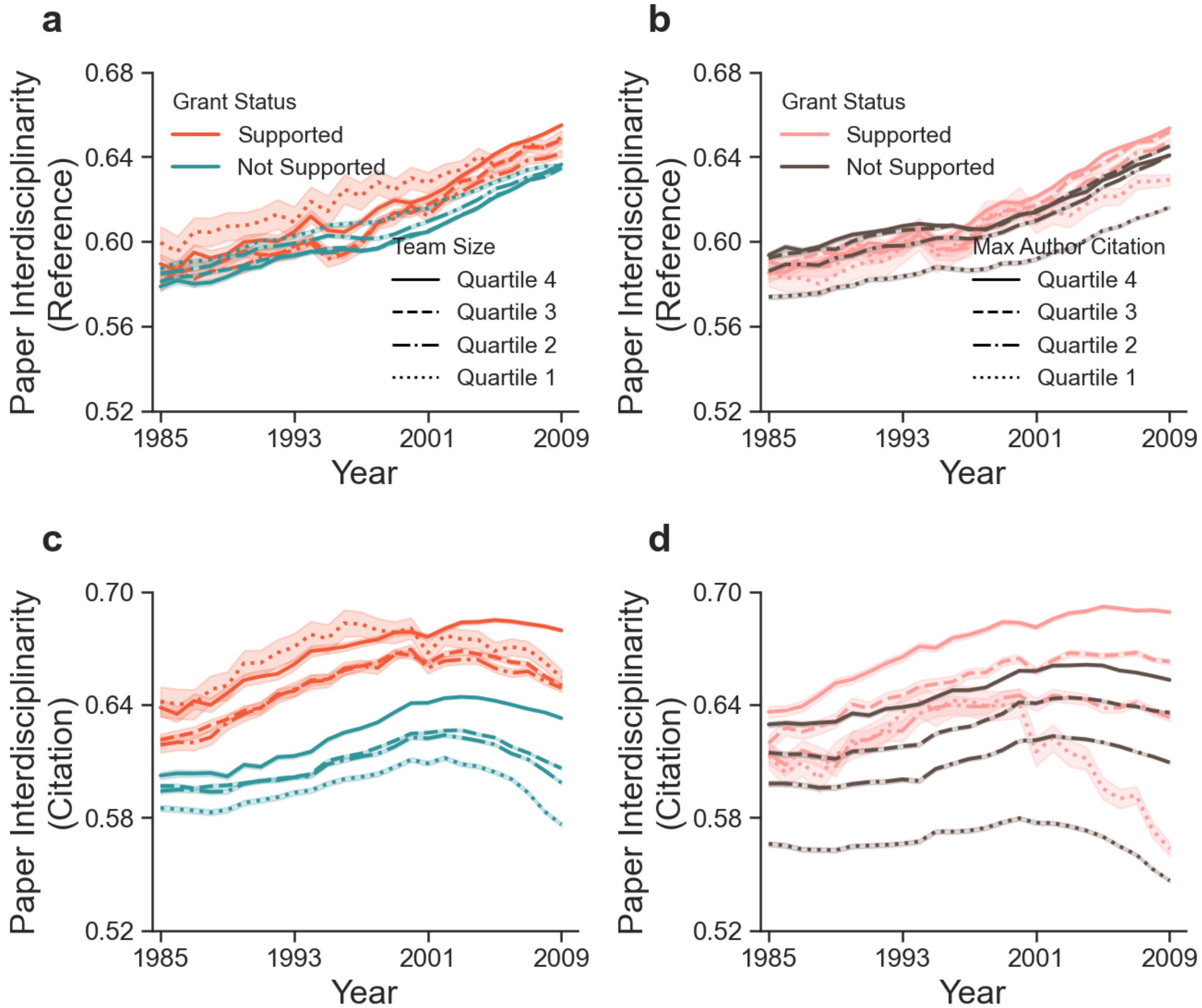

**Supplementary Fig. S3. Increased interdisciplinarity of grant supported papers, controlling for author prominence. a,** Papers acknowledging grant support (red lines) display a persistently higher reference-based interdisciplinarity compared to those without grant support (blue lines), across varying team sizes. **b,** This pattern is consistent when controlled for author prominence, measured by the highest citation counts among authors, where grant-supported papers (pink lines) maintain a higher interdisciplinarity than those without grant support (brown lines), especially when compared to papers of similar author prominence levels (same line types). **c** and **d,** Grant support is associated with increased citation-based interdisciplinarity, as well, with larger effects.

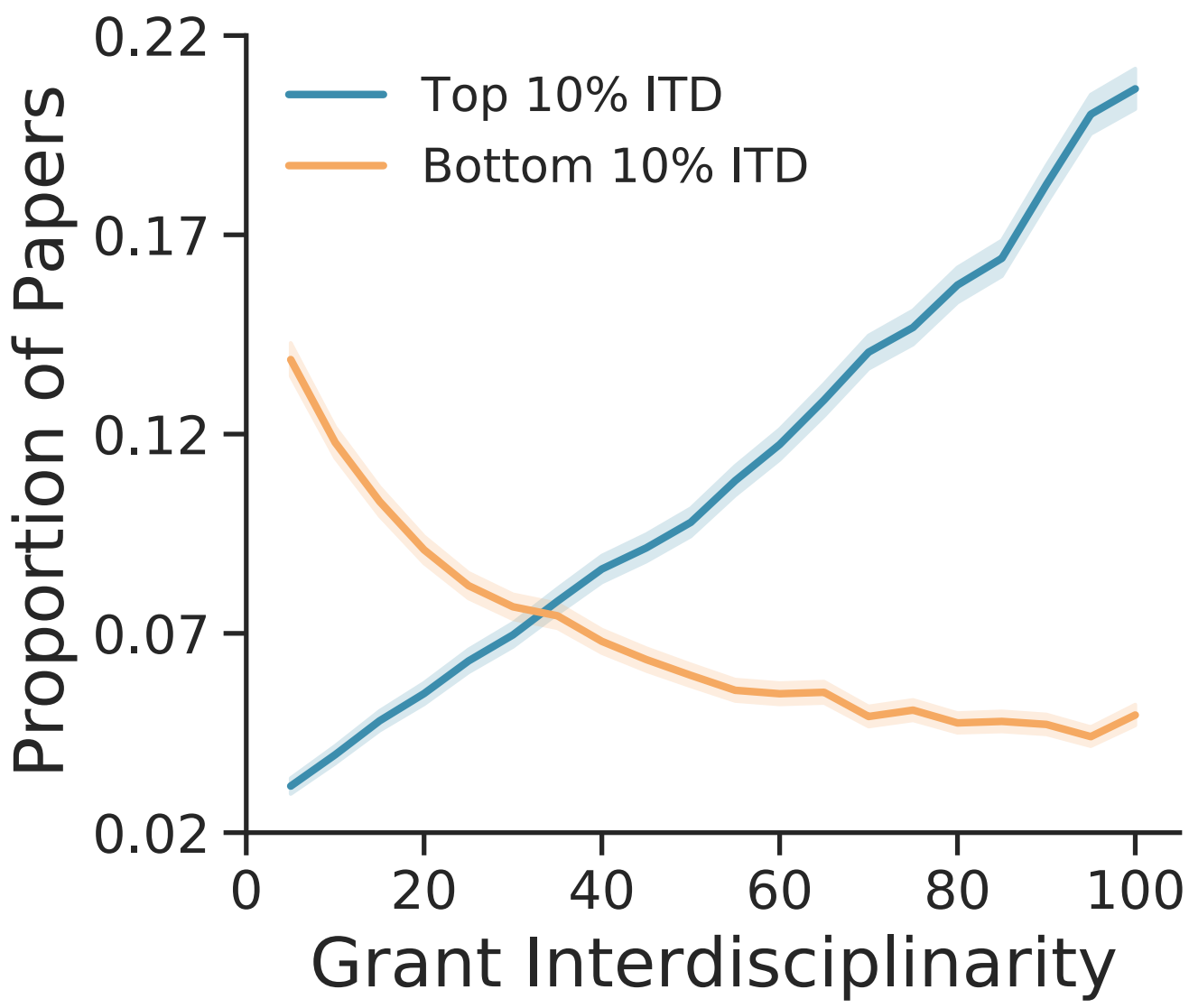

**Supplementary Fig. S4 | As expected, interdisciplinary grants were more likely to produce interdisciplinary papers while disciplinary grants tended to produce disciplinary papers.** Sorting publications according to their interdisciplinarity based on their references (ITD), we found that interdisciplinary grants tended to produce an increasing fraction of highly interdisciplinary papers (blue) that we defined as the top 10% of the papers ranked by interdisciplinarity. In turn, disciplinary grants supported more disciplinary publications (orange), which were defined as the bottom 10% of the papers ranked by interdisciplinarity.

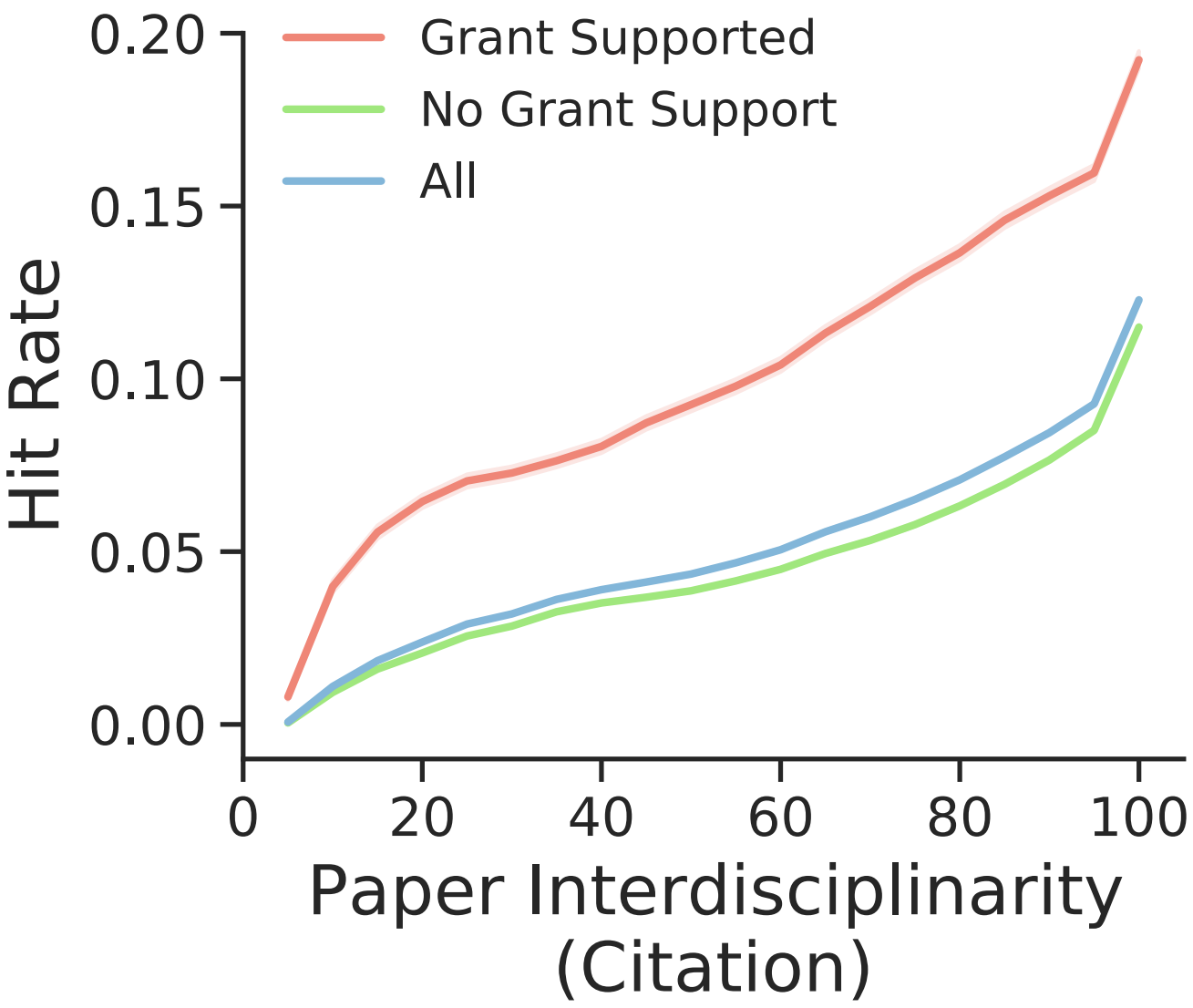

**Supplementary Fig. S5 | The impact of papers increased with the higher diversity of disciplines that the paper influenced.** The hit rate of papers is the probability that a paper appears in the top 5% in the field and year in terms of the number of citations. By measuring the hit rate of papers as a function of interdisciplinary appeal (i.e., citation-based paper interdisciplinarity), we find that more interdisciplinary papers have a greater impact. Furthermore, trends are enforced when we considered papers supported by grants.

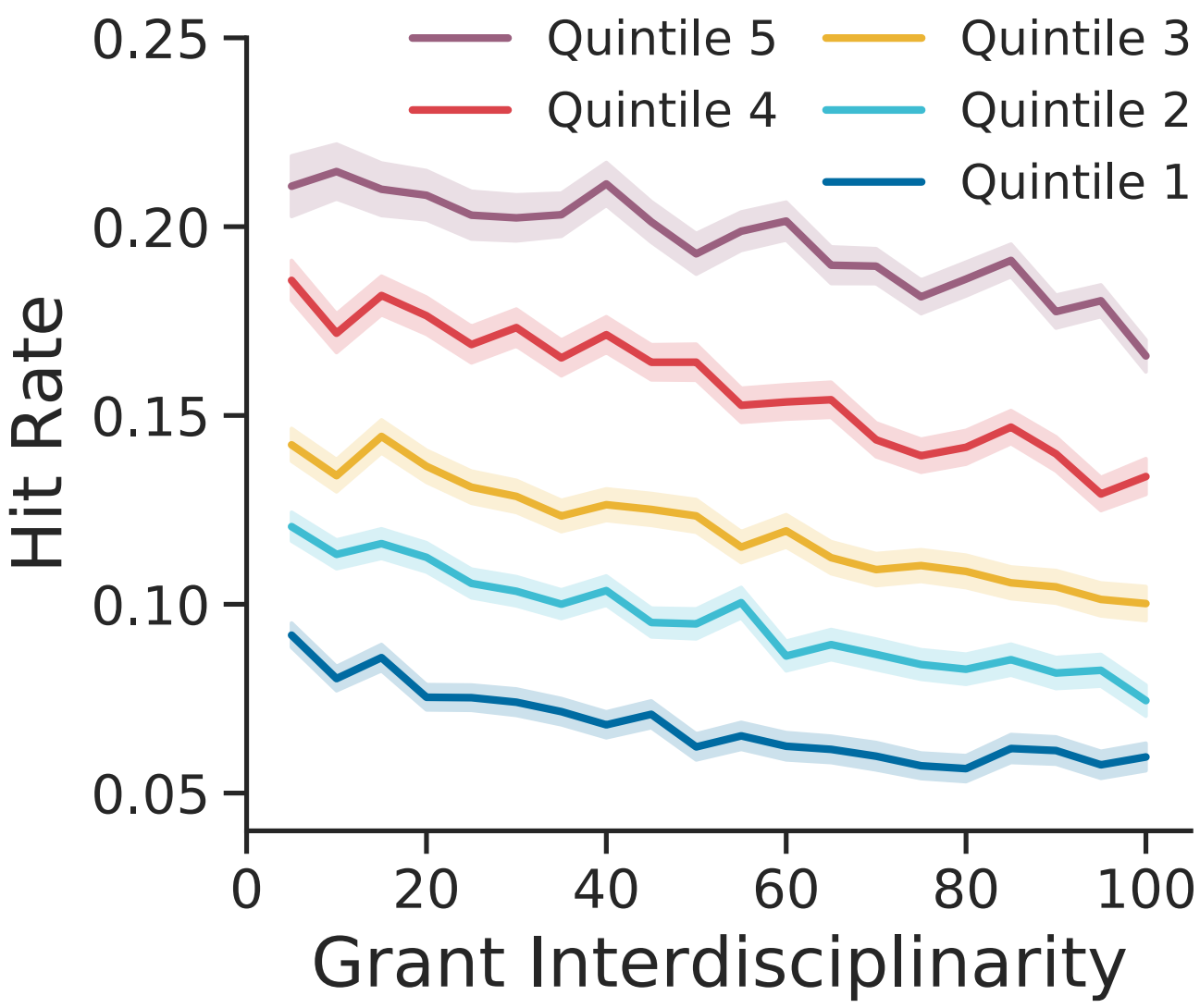

**Supplementary Fig. S6 | Interdisciplinary papers from more disciplinary grants tended to be associated with greater impact.** While interdisciplinary papers as a function of their citations had a greater chance of being hit papers (from Quintile 1 to Quintile 5), more disciplinary grants tended to support research with higher impact when comparing papers within the same interdisciplinarity level.

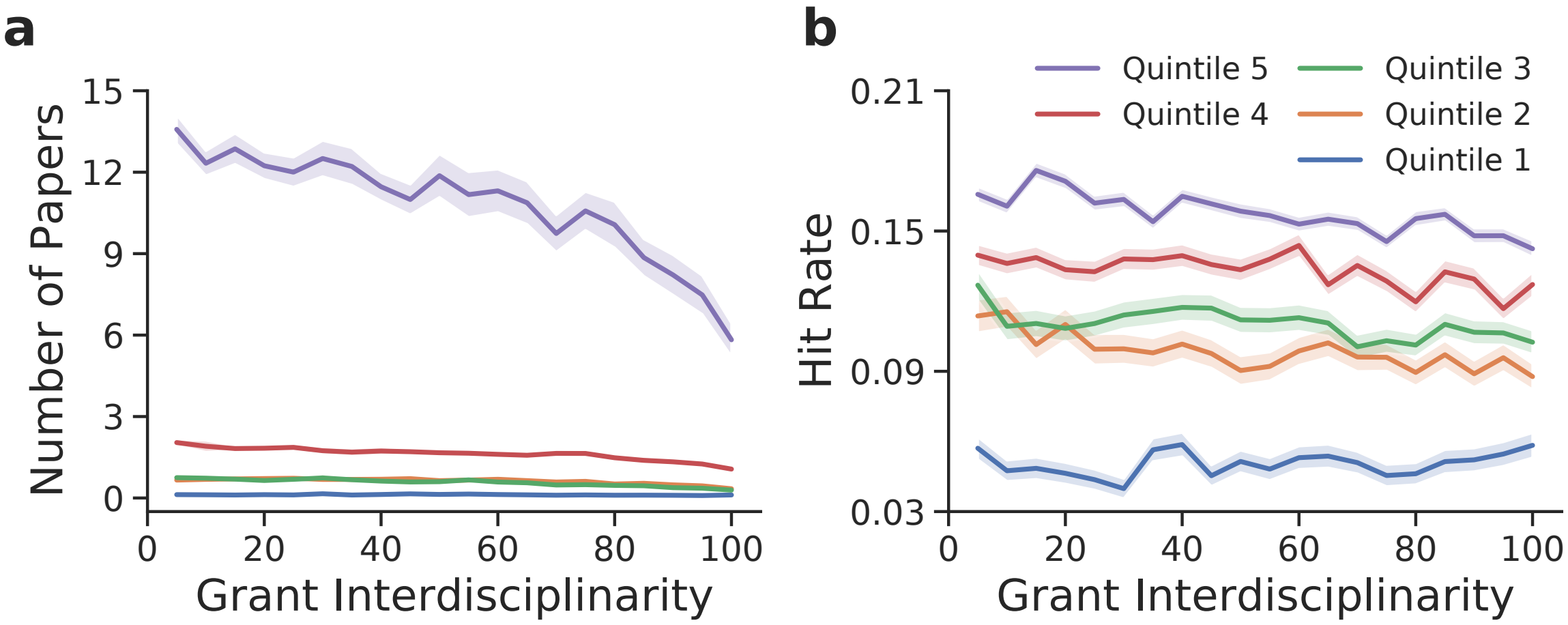

**Supplementary Fig. S7 | Increased publication productivity and impact of disciplinary grants at varied funding sizes. a,** Within each quintile of funding amounts, we find a trend of diminishing returns in paper production as grant interdisciplinarity grows. Notably, this effect is more pronounced for grants with larger funding amounts (from Quintile 1 to 5, ranging from smaller to larger funding amounts). Nevertheless, grants with larger budgets maintain a higher baseline for average publication numbers. **b,** Highly funded (Quintile 5), discipline-focused grants are most likely to yield hit papers. Across all but the lowest funding levels, there is a consistent rise in the average hit rate as grants become more discipline-specific, with this relationship strengthening in tandem with grant size.

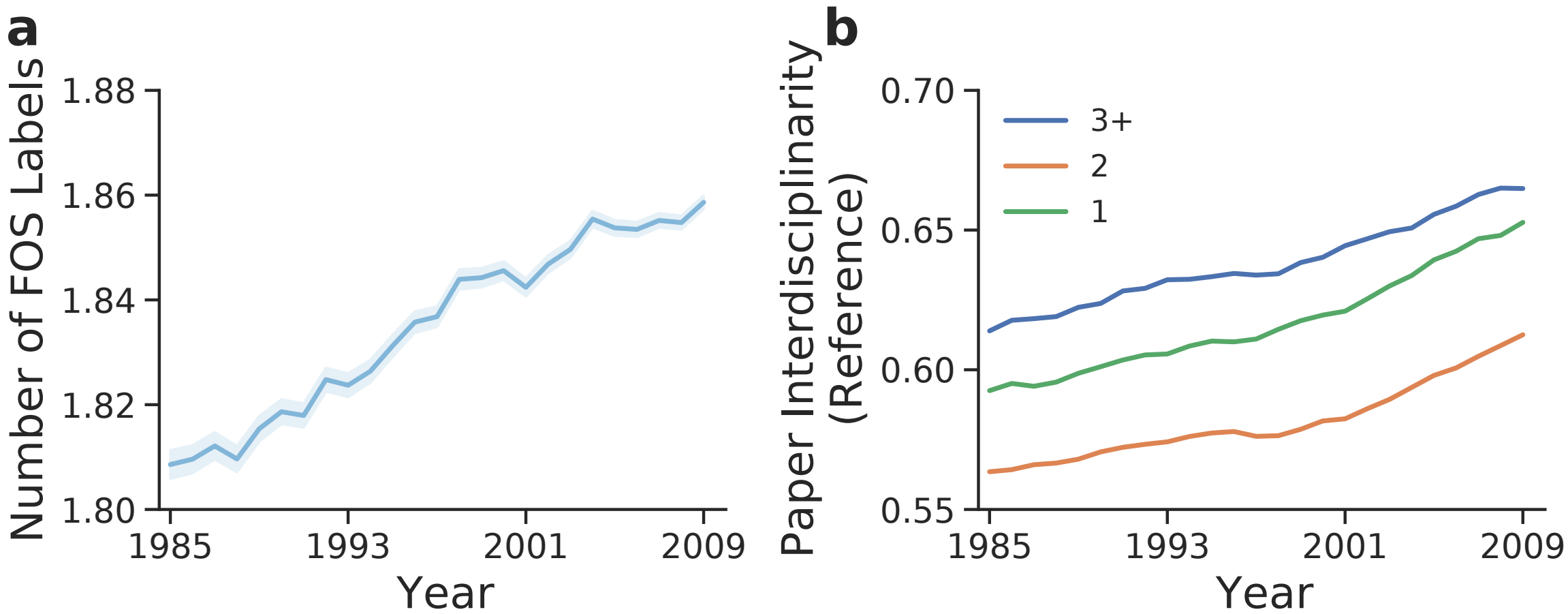

**Supplementary Fig. S8 | Consistent temporal increase in the interdisciplinarity of papers, observed even when conditioned on the number of Fields of Study (FOS) associated with each paper. a,** The average FOS labels per paper shows an incremental rise of around 3% from 1985 to 2009. **b,** The level of paper interdisciplinarity, as measured by references, increases from 1985 to 2009. The increasing trend in the interdisciplinarity is highly similar across papers with different numbers of associated FOS. However, papers associated with a greater number of FOS exhibit a higher baseline level of interdisciplinarity.

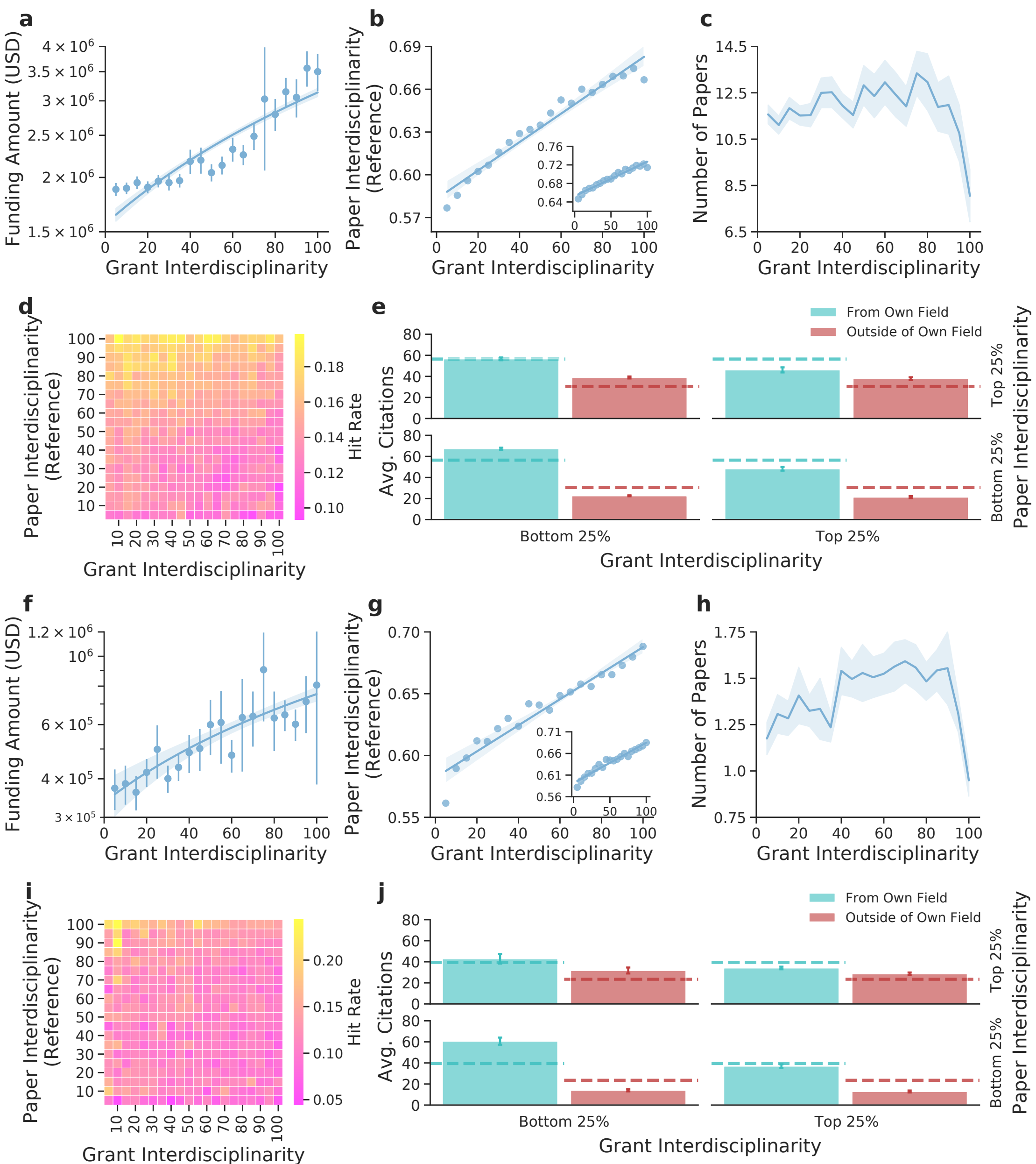

**Supplementary Fig. S9 | Robustness of main results across different major funding agencies, including NIH (a-e) and NSF (f-j).** Consistently, more interdisciplinary grants tend to secure larger funding amounts (**a**, **f**). The interdisciplinarity of papers, based on references (citations; inset), increases with the interdisciplinarity of the supporting grants (**b**, **g**). Highly interdisciplinary grants show a lower propensity to produce publications (**c**, **h**). Interdisciplinary papers supported by more disciplinary grants tend to garner higher impact (**d**, **i**). Interdisciplinary papers backed by disciplinary grants (top left) tend to receive similar or more citations than random baselines, both from within and outside their fields. In contrast, other types of papers attract comparable or more citations than the random baseline either from their own field (disciplinary papers supported by disciplinary grants; bottom left), from outside their field (interdisciplinary papers supported by interdisciplinary grants; top right), or neither (disciplinary papers supported by interdisciplinary grants; bottom right; **e**, **j**).

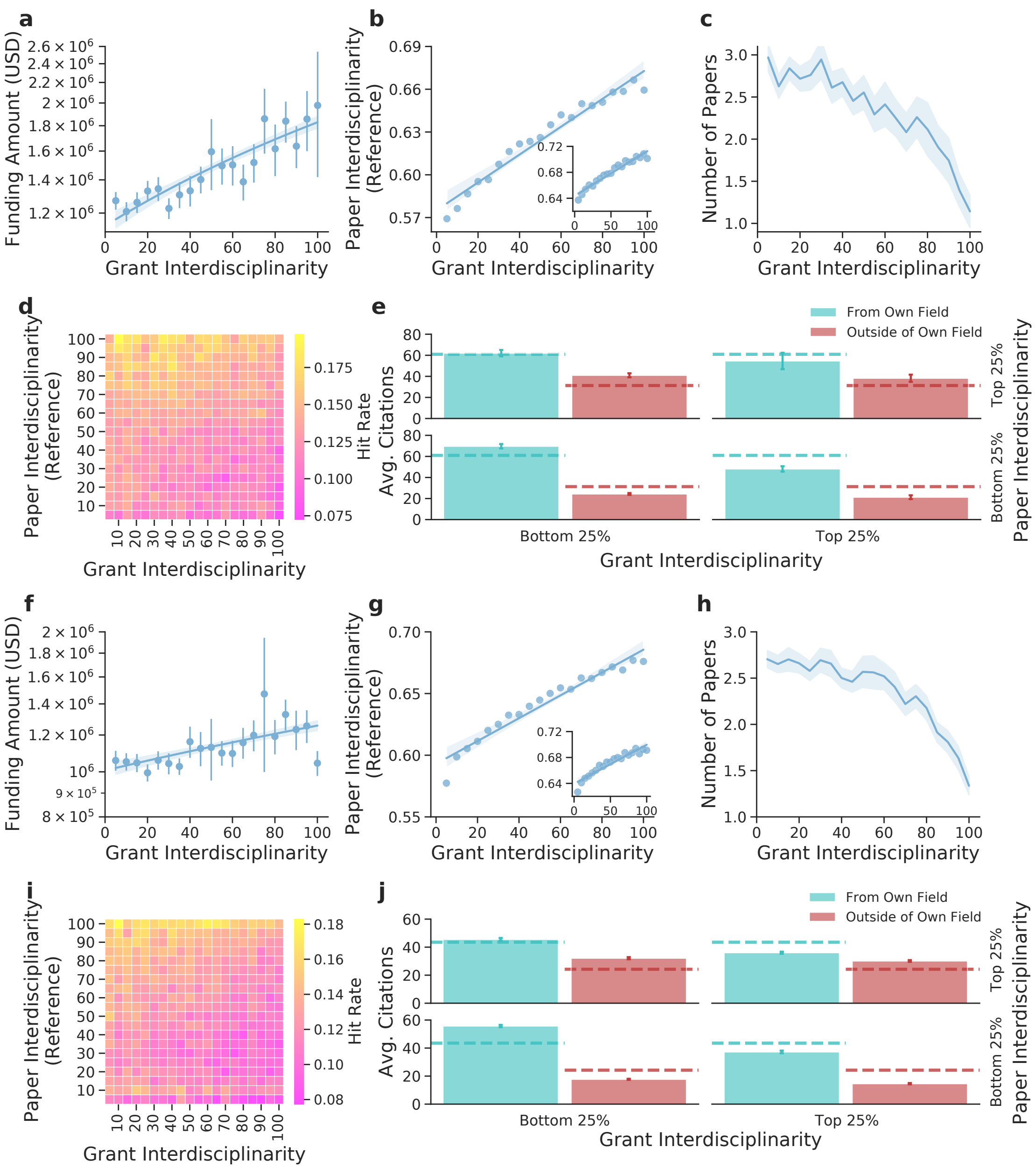

**Supplementary Fig. S10 | Robustness of main results across different time windows, before 2000 (a-e) and after 2000 (f-j).** Consistently, more interdisciplinary grants tend to secure larger funding amounts (**a**, **f**). The interdisciplinarity of papers, based on references (citations; inset), increases with the interdisciplinarity of the supporting grants (**b**, **g**). Highly interdisciplinary grants show a lower propensity to produce publications (**c**, **h**). Interdisciplinary papers supported by more disciplinary grants tend to garner higher impact (**d**, **i**). Interdisciplinary papers backed by disciplinary grants (top left) tend to receive similar or more citations than random baselines, both from within and outside their fields. In contrast, other types of papers attract comparable or more citations than the random baseline either from their own field (disciplinary papers supported by disciplinary grants; bottom left), from outside their field (interdisciplinary papers supported by interdisciplinary grants; top right), or neither (disciplinary papers supported by interdisciplinary grants; bottom right; **e**, **j**).

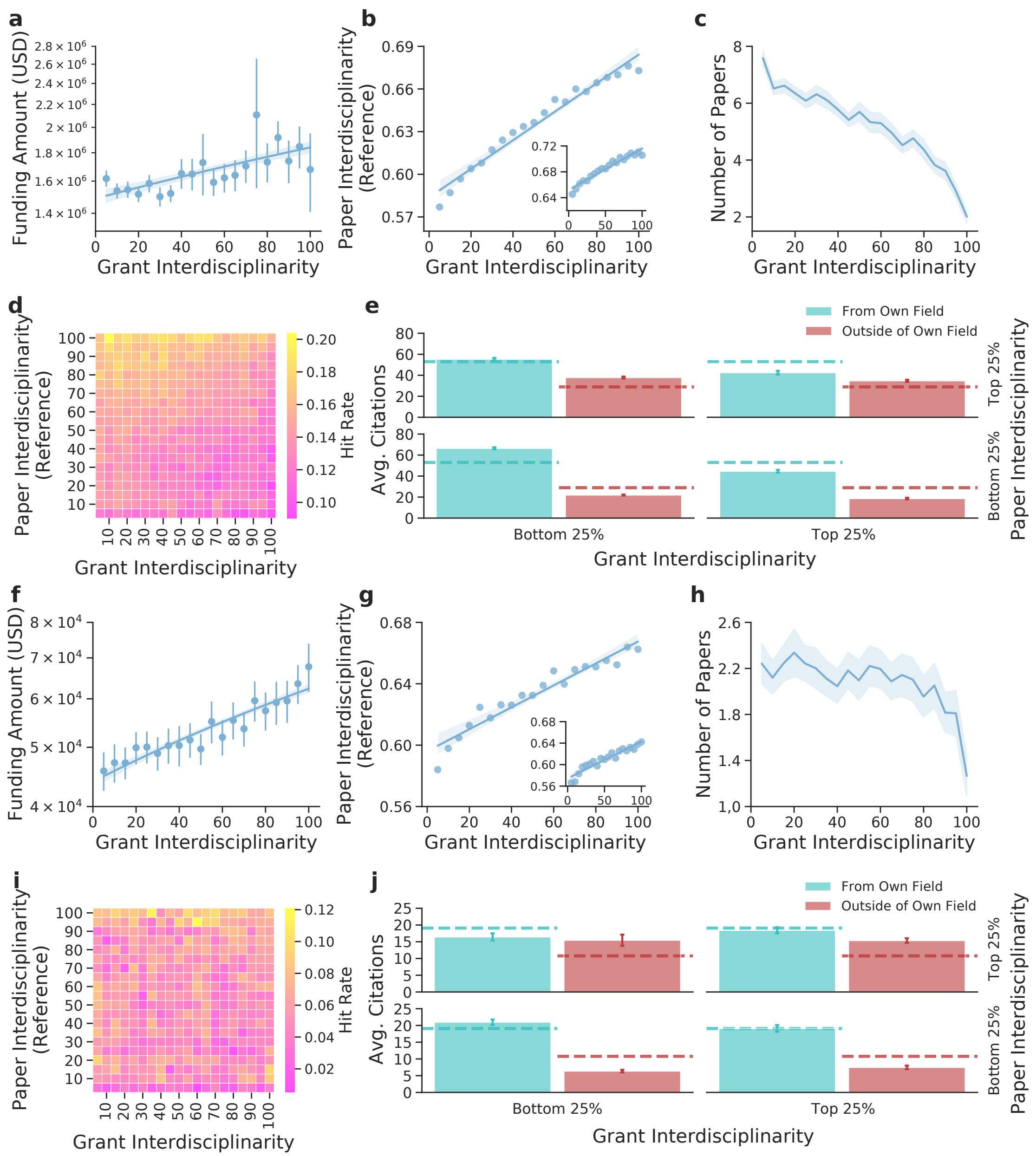

**Supplementary Fig. S11 | Robustness of main results across culturally different countries, including the United States (a-e) and China (f-j).** Consistently, more interdisciplinary grants tend to secure larger funding amounts (**a**, **f**). The interdisciplinarity of papers, based on references (citations; inset), increases with the interdisciplinarity of the supporting grants (**b**, **g**). Highly interdisciplinary grants show a lower propensity to produce publications (**c**, **h**). Interdisciplinary papers supported by more disciplinary grants tend to garner higher impact (**d**, **i**). Interdisciplinary papers backed by disciplinary grants (top left) tend to receive similar or more citations than random baselines, both from within and outside their fields. In contrast, other types of papers attract comparable or more citations than the random baseline either from their own field (disciplinary papers supported by disciplinary grants; bottom left), from outside their field (interdisciplinary papers supported by interdisciplinary grants; top right), or a mix (disciplinary papers supported by interdisciplinary grants; bottom right; **e**, **j**).

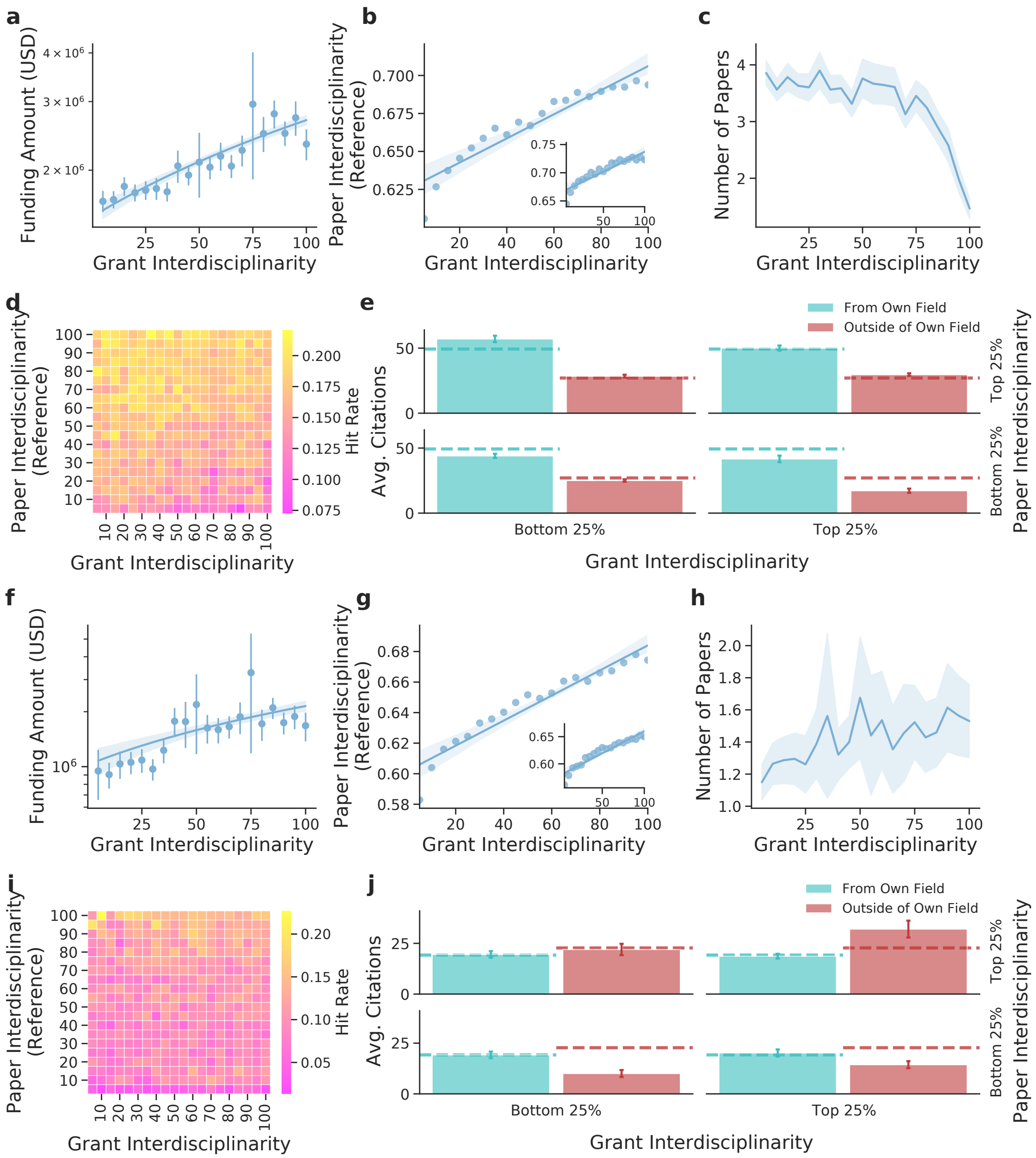

**Supplementary Fig. S12 | Robustness of main results across different disciplines, including Applied Sciences (a-e) and Formal Sciences (f-j).** Consistently, more interdisciplinary grants tend to secure larger funding amounts (**a**, **f**). The interdisciplinarity of papers, based on references (citations; inset), increases with the interdisciplinarity of the supporting grants (**b**, **g**). In Applied Sciences, highly interdisciplinary grants are less likely to produce publications (**c**), whereas more interdisciplinary grants tend to result in a higher number of papers in Formal Sciences (**h**). In both disciplines, papers that are interdisciplinary and supported by disciplinary grants tend to achieve higher impact (**d**, **i**). In both Applied Sciences and Formal Sciences, interdisciplinary papers backed by disciplinary grants tend to receive similar or higher citation counts than random baselines, both from within and outside their fields (top left in **e** and **j**). But, in Formal Sciences, highly interdisciplinary papers supported by highly interdisciplinary grants tend to receive significantly more citations from outside their core field, while maintaining expected citation levels from within the core field (top right in **j**).

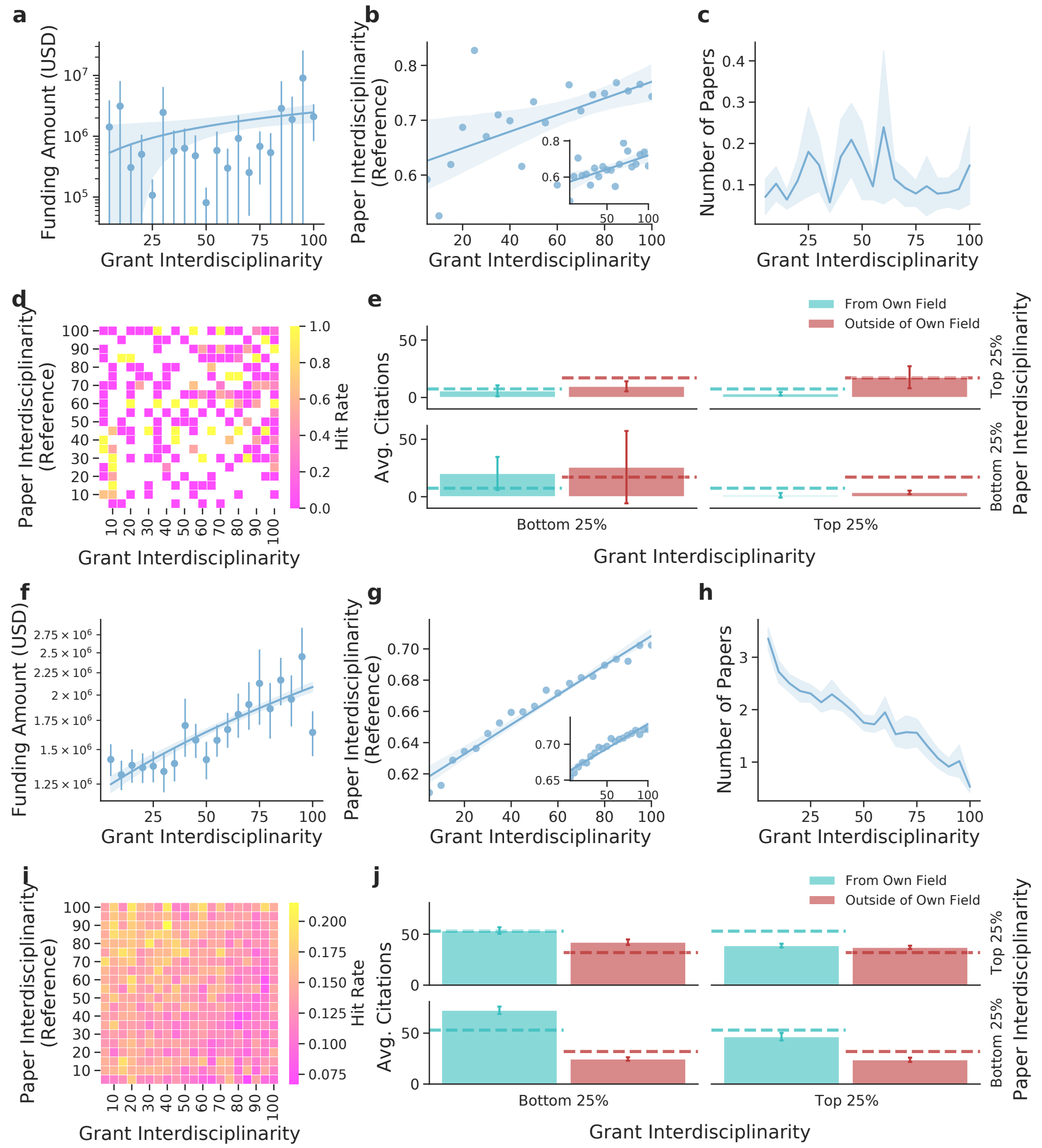

**Supplementary Fig. S13 | Robustness of main results across different disciplines, including Humanities (a-e) and Social Sciences (f-j).** In both Humanities and Social Sciences, grants with higher interdisciplinarity typically secured larger funding amounts (**a**, **f**). Paper interdisciplinarity based on references (citations; inset) increased as a function of grant interdisciplinarity (**b**, **g**). The number of papers supported by grants in Humanities plateaued with increasing grant interdisciplinarity (**c**), whereas a decreasing trend was observed in Social Sciences, similar to other conditions (**h**). The hit rate, conditional on both grant and paper interdisciplinarity, was less clear in Humanities due to limited data points (**d**), but, in Social Sciences, interdisciplinary papers supported by more disciplinary grants were associated with higher impact, aligning with trends in other conditions (**i**). In Humanities, highly disciplinary papers supported by disciplinary grants garnered more citations than expected from both within and outside their own field (bottom left in **e**). In contrast, this high and broad impact was observed for highly interdisciplinary papers supported by highly disciplinary grants in Social Sciences (top left in **j**). Note that the results pertaining to Humanities should be interpreted with caution due to the limited number of data points, which affects statistical reliability.

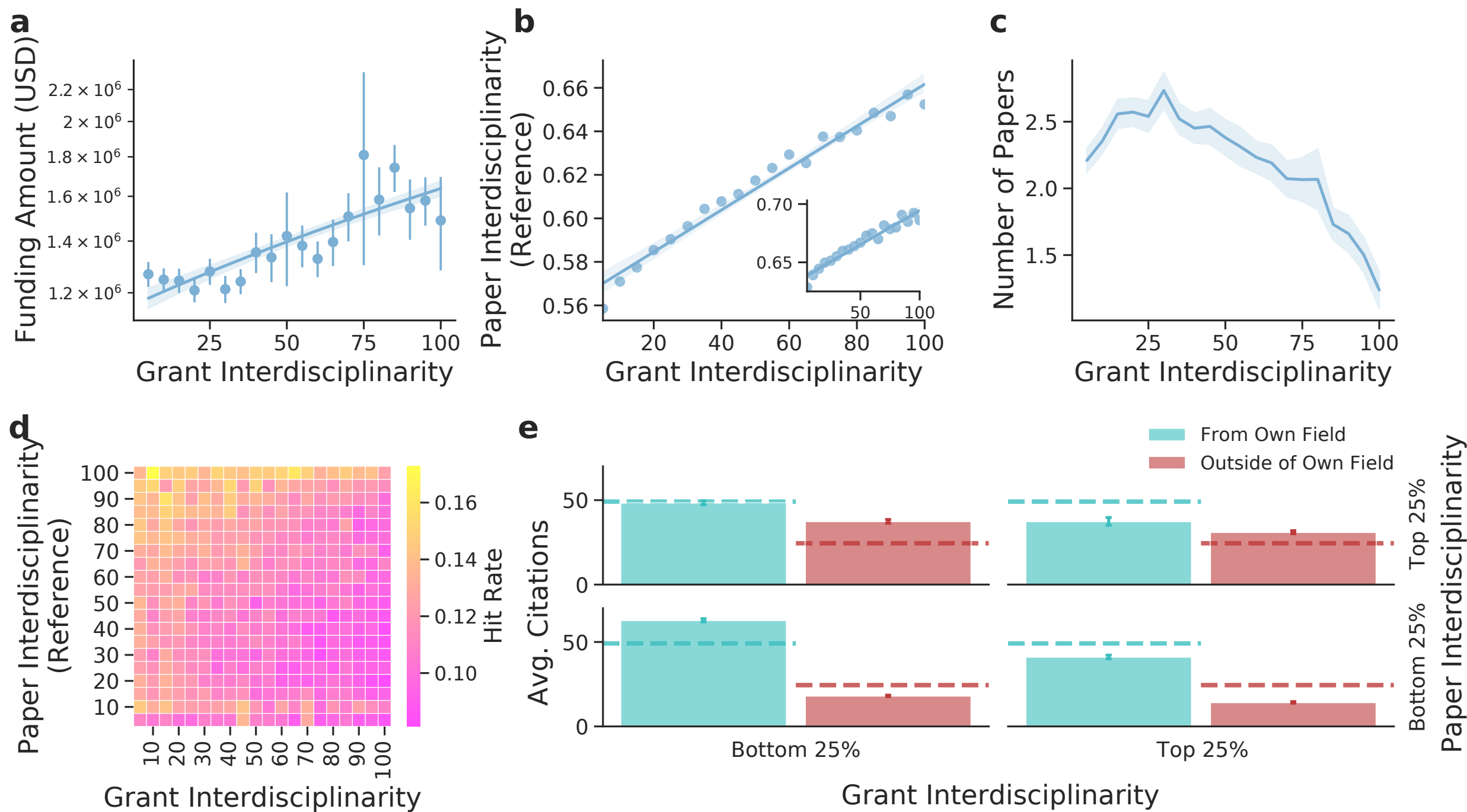

**Supplementary Fig. S14 | Robustness of main results in Natural Sciences.** Consistently, more interdisciplinary grants tend to secure larger funding amounts (**a**). The interdisciplinarity of papers, based on references (citations; inset), increases with the interdisciplinarity of the supporting grants (**b**). Highly interdisciplinary grants show a lower propensity to produce publications (**c**). Interdisciplinary papers supported by more disciplinary grants tend to garner higher impact (**d**). Interdisciplinary papers backed by disciplinary grants (top left; **e**) tend to receive similar or more citations than random baselines, both from within and outside their fields. In contrast, other types of papers attract comparable or more citations than the random baseline either from their own field (disciplinary papers supported by disciplinary grants; bottom left), from outside their field (interdisciplinary papers supported by interdisciplinary grants; top right), or neither (disciplinary papers supported by interdisciplinary grants; bottom right).

**Supplementary Table S1.** Fields with most representative words by probability and FREX score.

| Field | Weight Type | Top 10 Words |
|---|---|---|
| **Accounting** | **Probability** | account, financi, manag, corpor, use, compani, audit, studi, govern, report |
| | **FREX** | auditor, audit, ifr, accrual, disclosur, gaap, csr, cpa, ceo, sharehold |
| **Acoustics** | **Probability** | use, acoust, frequenc, measur, sound, nois, signal, result, method, wave |
| | **FREX** | acoust, transduc, microphon, masker, loudspeak, hydrophon, reverber, piezoelectr, interaur, sonar |
| **Actuarial Science** | **Probability** | insur, risk, use, model, cost, studi, health, financi, paper, rate |
| | **FREX** | insur, annuiti, hmo, actuari, medicar, qali, nonfinanci, enrolle, mco, fsa |
| **Advertising** | **Probability** | advertis, use, product, media, consum, studi, market, brand, sport, effect |
| | **FREX** | advertis, wom, volleybal, basketbal, championship, brand, garvey, televis, c2c, olymp |
| **Aeronautics** | **Probability** | aircraft, flight, develop, system, aviat, air, crew, mission, control, safeti |
| | **FREX** | visor, warhead, powerpl, airspac, aircrew, cross-kick, front-row, usafa.edu, gadss, ohb |
| **Aerospace Engineering** | **Probability** | use, flight, design, system, test, perform, result, model, aircraft, develop |
| | **FREX** | hyperson, scramjet, thruster, airship, airfram, flowfield, hover, rocket, railgun, cubelab |
| **Aesthetics** | **Probability** | cultur, aesthet, art, work, music, one, life, also, modern, form |
| | **FREX** | aesthet, beckett, beauti, kemal, kemalist, alaranta, skin-bleach, woodcock, yang-m, doll |
| **Agricultural Economics** | **Probability** | product, agricultur, use, farm, price, increas, land, food, develop, farmer |
| | **FREX** | acreag, smallhold, t+1, pakcoy, post-harvest, sokoto, basod, fuelwood, obr, mendong |
| **Agricultural Engineering** | **Probability** | crop, agricultur, water, develop, rice, irrig, system, use, method, model |
| | **FREX** | sprinkler, harrow, eucommia, wacm, non-cultiv, capacity-bas, scallion, htp, hill-drop, fertilizer-middl |
| **Agricultural Science** | **Probability** | product, farmer, farm, use, studi, agricultur, market, crop, produc, research |
| | **FREX** | a©, gapoktan, lkm-a, bugday, kvk, khat, komponen, produktivita, petani, sistim |
| **Agroforestry** | **Probability** | forest, use, speci, soil, tree, product, land, manag, area, plant |
| | **FREX** | agroforestri, plantat, shrub, woodi, savanna, grassland, understori, woodland, homegarden, rangeland |
| **Agronomy** | **Probability** | soil, yield, plant, increas, n, effect, crop, use, differ, fertil |
| | **FREX** | tiller, weed, tillag, ryegrass, mulch, manur, npk, sorghum, panicl, clover |
| **Algebra** | **Probability** | algebra, gener, group, paper, use, theori, result, function, represent, properti |
| | **FREX** | drinfeld, quasi-hopf, monoid, lusztig, groebner, bialgebra, morita, galoi, pbw, krasner |
| **Algorithm** | **Probability** | algorithm, use, method, propos, result, base, data, model, paper, perform |
| | **FREX** | ldpc, precod, doa, ofdm, step-siz, glrt, papr, decod, frequency-select, watermark |
| **Analytical Chemistry** | **Probability** | use, film, temperatur, measur, sampl, surfac, result, increas, studi, method |
| | **FREX** | sputter, anneal, magnetron, xp, undop, photoluminesc, voltammetri, photoelectron, dope, as-deposit |
| **Anatomy** | **Probability** | cell, muscl, studi, nerv, use, differ, result, develop, increas, neuron |
| | **FREX** | axon, innerv, ganglion, nerv, dorsal, immunoreact, caudal, ventral, cartilag, myelin |
| **Ancient History** | **Probability** | centuri, histori, dynasti, period, war, time, year, first, one, empir |
| | **FREX** | shang, emperor, ottoman, zoroastrian, seljuk, constantinopl, jori, haile-selassi, yohann, patani |
| **Andrology** | **Probability** | group, sperm, embryo, cell, oocyt, rate, use, day, fertil, studi |
| | **FREX** | spermatozoa, acrosom, sperm, blastocyst, semen, cryopreserv, vitrif, oocyt, frozen-thaw, vitrifi |
| **Anesthesia** | **Probability** | group, patient, effect, studi, use, blood, treatment, p, increas, control |
| | **FREX** | anesthesia, analgesia, propofol, anaesthesia, anesthet, epidur, bupivacain, fentanyl, lidocain, intub |
| **Animal Science** | **Probability** | p, group, weight, day, differ, diet, effect, increas, feed, use |
| | **FREX** | heifer, calv, ewe, carcass, wean, sire, crossbr, holstein, rumen, cow |
| **Anthropology** | **Probability** | cultur, studi, anthropolog, social, articl, peopl, histori, paper, research, polit |
| | **FREX** | shaman, anthropologist, saami, gvp, bugi, ethnolog, anthropolog, nuer, comodif, ethnograph |
| **Applied Mathematics** | **Probability** | method, model, equat, use, problem, solut, function, system, estim, numer |
| | **FREX** | pitd, b-poli, ode, divergence-clean, karhunen–loev, krylov, volterra, sode, md-lvq, hemivari |
| **Applied Psychology** | **Probability** | use, studi, research, job, train, work, perform, psycholog, result, effect |
| | **FREX** | coach, hockey, rdm, emoji, mouthguard, bulli, luckasson, shiftwork, workout, ebd |
| **Archaeology** | **Probability** | archaeolog, site, use, studi, cultur, area, date, new, one, period |
| | **FREX** | archaeolog, potteri, prehistor, neolith, archaeologist, bronz, figurin, sherd, palaeolith, obsidian |
| **Architectural Engineering** | **Probability** | build, design, energi, use, system, paper, space, studi, architectur, construct |
| | **FREX** | courtyard, hvac, glaze, bipv, leed, air-condit, occupants’, multi-famili, gshp, biophil |
| **Arithmetic** | **Probability** | arithmet, number, use, adder, comput, multipli, method, design, multipl, paper |
| | **FREX** | soal, h*-algebra, kisi-kisi, ohrn, place-valu, n-bit, clz, adder, m-spotti, jscac |
| **Art History** | **Probability** | work, de, art, new, one, book, first, year, time, histori |
| | **FREX** | rembrandt, abbott, painter, tarzan, terezin, nin, perrudja, welbi, matiss, kaempfer |
| **Artificial Intelligence** | **Probability** | use, model, system, network, learn, method, data, paper, propos, algorithm |
| | **FREX** | neural, cnn, backpropag, unsupervis, perceptron, mlp, dnn, robot, lstm, semi-supervis |

| **Astrobiology** | **Probability** | earth, mar, atmospher, surfac, impact, crater, planet, meteorit, solar, asteroid |
|---|---|---|
| | **FREX** | martian, meteorit, crater, asteroid, regolith, chondrit, astrobiolog, titan', shergottit, uranu |
| **Astronomy** | **Probability** | star, galaxi, observ, use, cluster, mass, system, present, result, stellar |
| | **FREX** | ngc, galaxi, photometri, stellar, photometr, dwarf, gyr, star, fe/h, mag |
| **Astrophysics** | **Probability** | observ, model, emiss, star, galaxi, line, x-ray, sourc, use, field |
| | **FREX** | pulsar, agn, grb, supernova, galact, quasar, accret, flare, redshift, halo |
| **Atmospheric Sciences** | **Probability** | model, observ, atmospher, measur, use, aerosol, cloud, temperatur, data, result |
| | **FREX** | stratospher, aerosol, tropospher, ozon, mesospher, microphys, thermospher, cirru, updraft, ppbv |
| **Atomic Physics** | **Probability** | energi, electron, state, use, ion, plasma, calcul, atom, result, measur |
| | **FREX** | rydberg, photoion, ioniz, cyclotron, autoion, kev, collision, auger, attosecond, vibron |
| **Audiology** | **Probability** | hear, patient, studi, use, test, result, subject, group, differ, auditori |
| | **FREX** | cochlear, tinnitu, hear, vestibular, nystagmu, sensorineur, audiometri, audiolog, abr, dpoae |
| **Automotive Engineering** | **Probability** | system, engin, vehicl, control, fuel, use, power, energi, electr, design |
| | **FREX** | brake, powertrain, egr, diesel, hev, gasolin, throttl, turbocharg, supercharg, phev |
| **Biochemical Engineering** | **Probability** | use, develop, chemic, method, bioreactor, system, engin, biolog, materi, cell |
| | **FREX** | kms005, c.robustum, mobili, microbiotest, bio-deriv, awc, efc, electroorgan, model-system, semi-mechanist |
| **Biochemistry** | **Probability** | activ, protein, acid, cell, enzym, effect, use, increas, studi, result |
| | **FREX** | microsom, reductas, pyruv, atpas, phospholipid, dehydrogenas, synthetas, phospholipas, purifi, kda |
| **Bioinformatics** | **Probability** | gene, use, protein, diseas, studi, develop, data, cancer, sequenc, method |
| | **FREX** | bioinformat, gwa, genome-wid, ptm, non-cod, mirna, pharmacogenom, lncrna, protocadherin, rna-seq |
| **Biological System** | **Probability** | model, predict, cell, network, data, paramet, protein, structur, quantit, develop |
| | **FREX** | smlm, time-aggreg, horse-tooth, d.e, bull', frap, mbei, rigescen, flow-ieg, tgt |
| **Biomedical Engineering** | **Probability** | use, tissu, cell, measur, studi, bone, method, imag, model, result |
| | **FREX** | scaffold, decellular, microneedl, tissue-engin, biomateri, biocompat, microbubbl, ivd, osteochondr, peuu |
| **Biophysics** | **Probability** | cell, membran, protein, use, channel, studi, activ, effect, structur, mechan |
| | **FREX** | ca2+, myosin, nucleosom, phospholamban, patch-clamp, pseudopod, protofibril, voltage-depend, psii, ncx |
| **Biotechnology** | **Probability** | product, use, plant, develop, food, genet, research, resist, method, new |
| | **FREX** | biotech, biotechnolog, germplasm, gmo, basmati, anuradhapura, bioprocess, single-cross, kirik, mycotoxin |
| **Botany** | **Probability** | plant, speci, use, differ, root, studi, growth, result, show, effect |
| | **FREX** | callu, auxin, plantlet, anther, phloem, cotyledon, xylem, mycorrhiz, explant, aphid |

| Business Administration | Probability | dan, yang, innov, ini, employe, research, der, develop, dalam, dengan |
|---|---|---|
| | FREX | personalo, ptsp, kjk, pemilihan, instansi, ka¶zvetlen, bisinosi, organisasi, pplh, rhodia |
| Calculus | Probability | problem, theori, mathemat, calculu, method, use, paper, function, chapter, develop |
| | FREX | yanduan, homeorhesi, lpa2v, veiculo, paraconsist, sea-wav, coimplic, semi-uninorm, micro-perfor, bigeometr |
| Cancer Research | Probability | cell, cancer, express, tumor, gene, activ, effect, protein, studi, inhibit |
| | FREX | pten, survivin, cyclin, p53, emt, p16, xenograft, hypermethyl, myc, pdac |
| Cardiology | Probability | patient, coronari, heart, arteri, ventricular, left, cardiac, p, myocardi, group |
| | FREX | ventricular, atrial, mitral, coronari, echocardiographi, myocardi, tachycardia, aortic, angina, echocardiograph |
| Cartography | Probability | map, de, use, spatial, area, studi, data, la, urban, model |
| | FREX | cartograph, cartographi, ið, denizli, við, troca, tað, linfoma, gebaud, cyberbulli |
| Cell Biology | Probability | cell, protein, activ, signal, express, regul, function, role, studi, pathway |
| | FREX | autophagi, cytoskeleton, microtubul, golgi, integrin, endosom, mitosi, gtpase, actin, endocytosi |
| Ceramic Materials | Probability | void/modulu, wang, pi=0.475, treatment."58, menella, si f, spin-hyperfin, spin-flavour, feminist/gend, triglyceride/cholesteryl |
| | FREX | void/modulu, pi=0.475, wang, amyloid-depend, acid-grown, drug-bear, melatonin-pretr, globulin-γ, lmrnol/l, channel-perm |
| Chemical Engineering | Probability | use, surfac, result, film, particl, structur, prepar, temperatur, show, properti |
| | FREX | tio2, calcin, mesopor, anatas, tio, mah, nanocomposit, titania, pani, nanospher |
| Chemical Physics | Probability | structur, surfac, dynam, molecul, molecular, model, studi, interact, use, simul |
| | FREX | h-bn, xe-129, nanopor, nanobubbl, dsdna, single-fil, nemat, thermophoresi, b-graphdiyn, counterion |
| Chromatography | Probability | use, method, extract, determin, sampl, concentr, acid, result, studi, detect |
| | FREX | chromatograph, chromatographi, hplc, elut, rsd, c18, reversed-phas, derivat, eluent, electrospray |
| Civil Engineering | Probability | construct, design, use, build, system, paper, structur, method, project, engin |
| | FREX | precast, formwork, pavement, asphalt, masonri, geotextil, semi-integr, geo-hazard, hma, dhw |
| Classical Economics | Probability | econom, growth, qualiti, russian, research, author, export, studi, structur, gener |
| | FREX | g-trust, russian, uber, decease", trusted', survey,2, trust.1, non-commod, coyl, savour |
| Classical Mechanics | Probability | equat, model, flow, use, wave, result, field, solut, effect, numer |
| | FREX | vortic, axisymmetr, incompress, reynold, vortex, streamwis, newtonian, inviscid, viscou, mech |
| Classics | Probability | histori, one, book, first, work, centuri, year, studi, time, univers |
| | FREX | herakl, andrew', lind, nereu, crapsey, wawruch, swett, haggadah, rita, powhatan |
| Climatology | Probability | model, climat, temperatur, chang, region, use, precipit, data, period, increas |
| | FREX | sst, enso, monsoon, interannu, westerli, reanalysi, eof, anticyclon, downscal, extratrop |
| **Clinical Psychology** | **Probability** | studi, use, depress, group, result, measur, symptom, scale, effect, associ |
| | **FREX** | ptsd, posttraumat, subscal, perfection, anxieti, nssi, adhd, alexithymia, beck, ideat |
| **Cognitive Psychology** | **Probability** | task, memori, process, studi, effect, experi, result, differ, use, cognit |
| | **FREX** | distractor, prefront, nonword, fmri, other-rac, metacognit, event-rel, aphas, recollect, precuneu |
| **Cognitive Science** | **Probability** | cognit, process, theori, develop, research, brain, human, system, concept, model |
| | **FREX** | self-knowledg, neurosci, barsal, mouse-track, neuroscientist, languag.-, sociocomplex, spivey, marr, mahasiswa |
| **Combinatorial Chemistry** | **Probability** | peptid, compound, use, synthesi, activ, librari, develop, select, new, drug |
| | **FREX** | flupep, thioester, ba-tpq, m6a, galectin-1, desthpdactylolid, ba-tpq-hydrogel, oeg, drug-lik, chemoinformat |
| **Combinatorics** | **Probability** | n, graph, g, k, x, number, set, p, f, r |
| | **FREX** | digraph, subgraph, undirect, vertex, matroid, hypergraph, polytop, cliqu, graph, n^ |
| **Commerce** | **Probability** | market, product, industri, develop, retail, consum, trade, competit, countri, good |
| | **FREX** | cashless, upholst, bitcoin, sofa, the , jewelleri, to , mt103, sc4, padano |
| **Communication** | **Probability** | experi, differ, visual, two, use, result, effect, present, respons, task |
| | **FREX** | saccad, playback, distractor, duckl, courtship, stroph, svv, basc, conspecif, fepc |
| **Composite Material** | **Probability** | use, composit, properti, materi, result, increas, mechan, temperatur, effect, surfac |
| | **FREX** | tensil, filler, epoxi, modulu, mortar, ceram, flexur, polypropylen, sinter, indent |
| **Computational Biology** | **Probability** | use, protein, gene, sequenc, genom, method, model, studi, structur, approach |
| | **FREX** | dpcr, conopeptid, pri-mirna, srna, tfbss, bcr-abl1, metaproteom, decon, crispr/cas9, proteom |
| **Computational Chemistry** | **Probability** | calcul, energi, use, structur, method, molecul, function, state, electron, bond |
| | **FREX** | initio, b3lyp, ccsd, mp2, hartree–fock, hyperpolariz, solvat, multirefer, chem, semiempir |
| **Computational Physics** | **Probability** | field, method, simul, use, calcul, model, particl, plasma, magnet, result |
| | **FREX** | zdr, altp, line-pair, vlf/lf, rbed, beh, kdp, zh, water–ic, cr39 |
| **Computational Science** | **Probability** | comput, use, simul, parallel, method, mesh, develop, algorithm, problem, grid |
| | **FREX** | mamico, thin-sheet, layer-pack, esfm, molecular-continuum, networks.-, cggverita, msmp, parallelis, meshfre |
| **Computer Architecture** | **Probability** | architectur, design, comput, system, perform, hardwar, use, applic, model, parallel |
| | **FREX** | sureselect, skx, soda-ii, mcsoc, peppher, ccga, vhdl-am, cross-cor, musra, subcachelin |
| **Computer Engineering** | **Probability** | design, comput, problem, model, process, system, data, applic, use, algorithm |
| | **FREX** | algor, mm-wave, hylcam, bg-gamp, tridaq, dcnn, nontermin, closest-vector, sub-6-ghz, efþcient |
| **Computer Graphics (Images)** | **Probability** | use, imag, model, graphic, data, display, visual, render, comput, system |
| | **FREX** | opengl, gamut, otogra, projector, stippl, shader, hologram, cd-atla, vtk, crossref |

| **Computer Hardware** | **Probability** | system, use, data, design, control, process, hardwar, implement, paper, signal |
|---|---|---|
| | **FREX** | overview.-, cpld, xilinx, mcu, chip.-, eeprom, a/d, micropost, fastbu, daq |
| **Computer Network** | **Probability** | network, propos, use, node, paper, perform, protocol, wireless, system, rout |
| | **FREX** | multicast, packet, manet, tcp, qo, handov, wsn, handoff, multi-hop, vanet |
| **Computer Security** | **Probability** | secur, system, attack, use, data, paper, inform, user, network, propos |
| | **FREX** | malici, password, malwar, authent, encrypt, ddo, cyber, signer, revoc, attack |
| **Computer Vision** | **Probability** | imag, use, method, propos, algorithm, result, object, base, detect, paper |
| | **FREX** | watermark, stereo, camera, pixel, hough, slam, jpeg, scene, rgb, registr |
| **Condensed Matter Physics** | **Probability** | magnet, temperatur, field, structur, effect, electron, result, state, phase, use |
| | **FREX** | ferromagnet, antiferromagnet, superconductor, phonon, magnetoresist, superconduct, josephson, kondo, superlattic, ferroelectr |
| **Construction Engineering** | **Probability** | construct, project, manag, design, engin, build, paper, inform, method, system |
| | **FREX** | fuze, civil-militari, ap1000, iptc, shipbreak, mine   select, lawnmow, sshac, self-seal, highway′ |
| **Control Engineering** | **Probability** | control, system, use, model, paper, design, power, method, robot, propos |
| | **FREX** | microgrid, servo, stator, droop, teleoper, brushless, pmsm, robot, mechatron, dfig |
| **Control Theory** | **Probability** | control, system, use, model, method, propos, paper, result, design, base |
| | **FREX** | pid, closed-loop, lyapunov, lmi, time-delay, feedforward, kalman, backstep, pwm, discrete-tim |
| **Criminology** | **Probability** | crime, crimin, violenc, polic, offend, social, victim, justic, studi, prison |
| | **FREX** | homicid, offend, crime, criminolog, gang, crimin, prison, offenc, recidiv, probat |
| **Crystallography** | **Probability** | structur, crystal, atom, two, phase, x-ray, c, complex, form, diffract |
| | **FREX** | orthorhomb, monoclin, single-cryst, triclin, tetragon, octahedr, tetrahedra, trigon, octahedra, unit-cel |
| **Data Mining** | **Probability** | data, use, method, model, algorithm, propos, base, result, system, paper |
| | **FREX** | itemset, skylin, outlier, apriori, k-mean, biclust, kdd, c4.5, top-k, e-contract |
| **Data Science** | **Probability** | data, research, use, inform, analysi, model, system, develop, paper, method |
| | **FREX** | predispens, vgi, bibliometr, cmda, aloja, cyberinfrastructur, epigenom, sherborn’, pridal, datam |
| **Database** | **Probability** | data, databas, system, use, inform, applic, manag, develop, paper, queri |
| | **FREX** | sql, hadoop, ldap, oracl, olap, warehous, databas, mysql, postgresql, hsct |
| **Demographic Economics** | **Probability** | incom, educ, inequ, household, effect, increas, countri, rate, women, growth |
| | **FREX** | de-industri, hukou, bribe, nonmetropolitan, in-migr, heirs’, efu, k12, deconcentr, otl |
| **Demography** | **Probability** | age, studi, year, use, rate, women, popul, among, mortal, risk |
| | **FREX** | non-hispan, menarch, age-adjust, condom, age-specif, skinfold, lbw, bmi, breakfast, overweight |

| Dentistry | Probability | group, use, studi, dental, teeth, bone, result, patient, implant, treatment |
|---|---|---|
| | FREX | periodont, dentin, cari, gingiv, teeth, dentur, endodont, enamel, tooth, edentul |
| Dermatology | Probability | patient, skin, treatment, case, diseas, clinic, report, lesion, use, studi |
| | FREX | psoriasi, erythema, dermat, dermatolog, dermatologist, acn, tinea, papul, alopecia, nevu |
| Development Economics | Probability | econom, countri, develop, polit, state, polici, region, social, nation, govern |
| | FREX | asean, zanzibar, neopatrimoni, hiv\|aid, anti-american, authoritarian, dhow, burundi, macapag, geopolit |
| Developmental Psychology | Probability | children, studi, differ, use, behavior, result, parent, age, group, test |
| | FREX | autism, iq, asd, preschool, prosoci, subtest, toddler, stutter, parent-child, adhd |
| Discrete Mathematics | Probability | set, function, gener, problem, use, result, n, algorithm, paper, show |
| | FREX | boolean, codeword, nondeterminist, automaton, polynomial-tim, submodular, t-norm, undecid, non-mal, pushdown |
| Distributed Computing | Probability | system, network, use, model, distribut, applic, comput, propos, paper, servic |
| | FREX | p2p, middlewar, replica, peer-to-p, datacent, qo, deadlock, self-stabil, sdn, fat-tre |
| Earth Science | Probability | earth, geolog, studi, use, scienc, geotherm, area, climat, ocean, isotop |
| | FREX | micro-xrf, rsl, kunlun, ygrc, nsb, geopp, garzanti, bruneau-grand, fine-s, weiser |
| Ecology | Probability | speci, popul, use, studi, differ, result, increas, effect, habitat, area |
| | FREX | habitat, predat, prey, herbivor, forag, parasitoid, brood, trophic, microhabitat, macroinvertebr |
| Econometrics | Probability | model, use, estim, data, result, method, paper, price, studi, test |
| | FREX | copula, garch, arima, out-of-sampl, autoregress, heteroskedast, semiparametr, cointegr, econometr, heteroscedast |
| Economic Geography | Probability | urban, citi, spatial, industri, region, econom, develop, agglomer, structur, growth |
| | FREX | agroecolog, industrial/sector, man-land, lan-xin, border-region, xi′an, homeplace-bas, cmrg′, gprd, wef |
| Economic Growth | Probability | develop, health, educ, countri, econom, social, rural, polici, govern, system |
| | FREX | mdg, microfin, rural, poverti, peasant, empower, unicef, countrysid, hiv/aid, livelihood |
| Economic History | Probability | war, polit, new, nation, histori, state, revolut, centuri, year, world |
| | FREX | stalin, lula, petrograd, unita, tonghak, shikai, i939, bolshevik, viet, lietuvininkai |
| Economic Policy | Probability | polici, econom, countri, govern, tax, fiscal, develop, reform, financi, public |
| | FREX | ceec, waemu, anti-money, bailout, re-elect, mdb, emtr, paygo, unibi, extra-budgetari |
| Economic System | Probability | develop, econom, economi, industri, system, social, region, polici, countri, govern |
| | FREX | soe, internationalis, post-socialist, subnat, inoguchi, foreign-invest, tnc, self-innov, europeanis, eoi |
| Economy | Probability | develop, econom, economi, region, industri, countri, paper, new, citi, market |
| | FREX | croissanc, zenmai, reunif, malaya, non-credit, wuppert, yodo, wine-mak, kib, mercosur |
| **Electrical Engineering** | **Probability** | power, system, use, voltag, current, circuit, design, paper, oper, high |
| | **FREX** | capacitor, breaker, igbt, inductor, kv, voltag, thyristor, overvoltag, high-voltag, charger |
| **Electronic Engineering** | **Probability** | use, system, design, propos, result, power, paper, signal, perform, present |
| | **FREX** | cmo, ofdm, antenna, microstrip, uwb, ghz, wideband, vco, demodul, bandpass |
| **Embedded System** | **Probability** | system, design, use, control, paper, applic, softwar, base, data, test |
| | **FREX** | zigbe, microcontrol, bluetooth, s-box, mpsoc, usb, ethernet, arm9, puf, mcu |
| **Emergency Medicine** | **Probability** | patient, hospit, care, use, studi, ed, emerg, result, medic, rate |
| | **FREX** | readmiss, ed, pddi, triag, delirium, in-hospit, stemi, triss, dvt, micu |
| **Endocrinology** | **Probability** | rat, increas, effect, cell, activ, level, express, receptor, studi, respons |
| | **FREX** | angiotensin, melatonin, acth, ang, adren, hypothalam, prolactin, corticosteron, hypothalamu, pituitari |
| **Engineering Drawing** | **Probability** | design, use, system, process, method, develop, machin, model, paper, part |
| | **FREX** | pro/e, lath, autocad, knit, solidwork, fixtur, cad/cam, cnc, pro/toolkit, cutting-stock |
| **Engineering Ethics** | **Probability** | research, educ, develop, scienc, ethic, practic, technolog, engin, scientif, new |
| | **FREX** | jmd, mfrc, strengths-bas, ebm, magdi, qir, backsourc, quality-ori, hta, wil |
| **Engineering Management** | **Probability** | manag, system, develop, project, engin, teach, technolog, paper, educ, design |
| | **FREX** | bcit, school-enterpris, comptia, cloudsm, iso9000, ssme, cdio, risk-inform, aiello, ganesha |
| **Engineering Physics** | **Probability** | materi, student, univers, engin, technolog, scienc, chemistri, research, energi, physic |
| | **FREX** | nbti, microvia, ipvt, hsinchu, screenprint, ibad, nemfet, solexel, cu3vo4, lmro |
| **Environmental Chemistry** | **Probability** | concentr, soil, water, use, organ, studi, sampl, metal, sediment, result |
| | **FREX** | pah, pbde, bioaccumul, ng/g, polychlorin, congen, pcdd/f, humic, pcb, mehg |
| **Environmental Economics** | **Probability** | energi, electr, system, develop, use, power, environment, econom, effici, model |
| | **FREX** | gscm, self-consumpt, feed-in, upss, ccgt, indc, pev, ricoh, v2g, growthfad |
| **Environmental Engineering** | **Probability** | water, use, concentr, studi, pollut, system, result, model, qualiti, emiss |
| | **FREX** | pm10, wwtp, coliform, denitrif, pm2.5, influent, biofilt, effluent, landfil, aerat |
| **Environmental Ethics** | **Probability** | human, cultur, develop, natur, social, societi, environment, ethic, peopl, life |
| | **FREX** | kinabalu, samskara, anthropocen, de-extinct, spondyloarthr, ecofeminist, samskaraâ€™, bajau, gada, flee |
| **Environmental Health** | **Probability** | health, use, studi, risk, exposur, among, result, diseas, associ, data |
| | **FREX** | asbesto, malaria, smokeless, smoke-fre, idu, stunt, farmwork, snack, tobacco, smoke |
| **Environmental Planning** | **Probability** | develop, urban, plan, environment, water, manag, area, use, land, citi |
| | **FREX** | resettl, eia, waterfront, brownfield, greenway, shadegan, sainj, parbati, city', land-use/cov |
| **Environmental Protection** | **Probability** | water, area, develop, environment, use, region, land, pollut, studi, protect |
| | **FREX** | icbp, bhutan, stockout, loch, possum, redd, tokai, tungiasi, geopark, wtr |
| **Environmental Resource Management** | **Probability** | develop, use, manag, environment, chang, sustain, system, resourc, studi, ecolog |
| | **FREX** | ecosystem, redd+, biodivers, social-ecolog, ecolog, emergi, esv, iczm, eco-econom, resili |
| **Epistemology** | **Probability** | theori, one, concept, natur, human, scienc, philosophi, paper, social, develop |
| | **FREX** | metaphys, kant, epistem, epistemolog, hegel, heidegg, philosoph, hume, nietzsch, husserl |
| **Ethnology** | **Probability** | de, le, et, la, cultur, histori, peopl, dan, nation, du |
| | **FREX** | mijikenda, haida, saramaka, beriberi, acadian, afro-hispan, anne, canadien, biafra, aiy |
| **Evolutionary Biology** | **Probability** | genet, speci, popul, evolut, select, gene, evolutionari, use, studi, differ |
| | **FREX** | phylogeni, supertre, cheater, half-chromatid, eutherian, neandert, trpr, simulan, angraecum, evol |
| **Family Medicine** | **Probability** | patient, health, care, use, studi, medic, clinic, result, practic, provid |
| | **FREX** | pharmacist, pharmaci, physician, pediatrician, condom, dietitian, prep, dentist, std, fgm |
| **Finance** | **Probability** | financi, bank, market, financ, invest, capit, use, manag, risk, firm |
| | **FREX** | financ, ipo, mortgag, underwrit, buyback, loan, cash, investor, ventur, estat |
| **Financial Economics** | **Probability** | market, price, stock, return, model, use, risk, result, volatil, trade |
| | **FREX** | arbitrag, hedg, dividend, reit, nyse, mean-vari, portfolio, capm, illiquid, cdo |
| **Financial System** | **Probability** | bank, financi, market, credit, risk, loan, system, crisi, sector, develop |
| | **FREX** | launder, npa, bank', payout, sbi, non-perform, inflasi, asset-bas, credit-spread, nonperform |
| **Fishery** | **Probability** | fish, speci, water, use, fisheri, studi, growth, differ, sea, rate |
| | **FREX** | spawn, salmon, fisheri, trout, hatcheri, crayfish, prawn, scallop, her, eel |
| **Food Science** | **Probability** | acid, use, product, content, studi, effect, increas, result, food, differ |
| | **FREX** | chees, juic, flour, whey, sausag, aroma, ferment, yogurt, dough, ddg |
| **Forensic Engineering** | **Probability** | use, accid, fire, paper, design, structur, method, concret, develop, caus |
| | **FREX** | flashov, smolder, picklex, stem–cement, lime-soil, windscreen, bloodstain, mbi, dnatypertm15, bomblet |
| **Forestry** | **Probability** | de, forest, area, tree, la, use, stand, le, studi, speci |
| | **FREX** | jalur, số, huevo, amenaza, ind./hm2in, larven, grain/m2, học, oncophora, larv |
| **Gastroenterology** | **Probability** | patient, group, treatment, liver, diseas, case, effect, rate, p, studi |
| | **FREX** | cirrhosi, pylori, gastriti, coliti, peptic, gerd, duoden, helicobact, omeprazol, dyspepsia |
| **Gender Studies** | **Probability** | women, cultur, gender, social, studi, ident, work, articl, sexual, polit |
| | **FREX** | feminist, masculin, femin, queer, lesbian, gay, transgend, lgbt, patriarchi, heterosexu |

| Genealogy | Probability | famili, name, time, histor, new, gener, year, present, one, histori |
|---|---|---|
| | FREX | cayuga, jungermannia, forfar, surnam, yup'ik, speck, tlingit, tik-tsam-sia, maji, sinodont |
| General Surgery | Probability | patient, surgeri, cancer, case, laparoscop, surgic, studi, oper, resect, perform |
| | FREX | laparoscop, hernia, cholecystectomi, gastrectomi, esophagectomi, laparotomi, anastomot, colostomi, laparoscopi, colectomi |
| Genetics | Probability | gene, sequenc, mutat, genet, genom, use, studi, dna, chromosom, region |
| | FREX | allel, chromosom, loci, exon, qtl, intron, haplotyp, codon, locu, telomer |
| Geochemistry | Probability | rock, deposit, miner, age, composit, isotop, melt, zone, element, magma |
| | FREX | plagioclas, granitoid, pluton, gneiss, zircon, mafic, clinopyroxen, xenolith, porphyri, granit |
| Geodesy | Probability | use, data, model, graviti, result, observ, gp, orbit, determin, satellit |
| | FREX | geoid, vlbi, goce, geodet, itrf, geopotenti, geodesi, dcb, insar, wgs-84 |
| Geometry | Probability | surfac, use, method, model, geometri, point, result, curv, flow, two |
| | FREX | precut, tα′, r~, sg20, microgroov, quasi-b-splin, hex-domin, efpim, vgtv, nonagon |
| Geomorphology | Probability | sediment, deposit, basin, area, structur, fault, result, rock, studi, region |
| | FREX | turbidit, morain, fluvial, glacier, prograd, foreland, tephra, dune, erosion, lacustrin |
| Geophysics | Probability | model, field, observ, wave, magnet, data, mantl, use, region, result |
| | FREX | substorm, auror, magnetospher, mantl, lithospher, magnetopaus, magnetotail, geomagnet, magnetosheath, daysid |
| Geotechnical Engineering | Probability | use, soil, model, test, result, rock, method, stress, effect, studi |
| | FREX | pile, grout, embank, triaxial, geotechn, subgrad, asphalt, seepag, undrain, scour |
| Gerontology | Probability | health, studi, age, use, activ, year, older, physic, associ, particip |
| | FREX | frailti, community-dwel, dementia, caregiv, sarcopenia, frail, sedentari, geriatr, gerontolog, adl |
| Gynecology | Probability | de, women, patient, cancer, use, studi, la, group, result, rate |
| | FREX | pacient, hpv, patienten, viaskin, endometri, iud, clomiphen, colposcopi, tratamiento, progestogen |
| Horticulture | Probability | fruit, plant, seed, effect, differ, increas, content, treatment, growth, result |
| | FREX | strawberri, ga3, bg, gourd, vine, rootstock, pusa, corm, uniconazol, postharvest |
| Humanities | Probability | de, la, en, que, el, e, lo, se, da, del |
| | FREX | relacion, educacion, mujer, aprendizaj, educativa, trabajo, texto, proceso, conocimiento, articulo |
| Human–Computer Interaction | Probability | user, use, system, design, interact, interfac, paper, develop, present, visual |
| | FREX | hci, human-comput, human-robot, usabl, sonif, gestur, multi-touch, tabletop, human-human, multitouch |
| Hydrology | Probability | water, model, use, river, area, flow, studi, soil, result, data |
| | FREX | runoff, catchment, hydrolog, aquif, recharg, groundwat, watersh, streamflow, evapotranspir, floodplain |

| Immunology | Probability | cell, patient, respons, studi, activ, immun, diseas, express, use, result |
|---|---|---|
| | FREX | nk, cd4+, treg, gvhd, cytokin, ige, lymphocyt, autoimmun, cd8, t-cell |
| Industrial Engineering | Probability | system, model, product, process, time, use, control, paper, oper, optim |
| | FREX | wlm, milk-run, sm-cc, windpow, moneymak, time-disturb, strali, a.d.hal, pert/cpm, prox-funct |
| Industrial Organization | Probability | industri, develop, market, firm, product, competit, innov, technolog, enterpris, paper |
| | FREX | subsidiari, tanzanit, sericultur, nev, aftermarket, msme, fabless, small-medium, decisions.-, shocks.- |
| Information Retrieval | Probability | use, inform, queri, document, retriev, user, search, data, semant, system |
| | FREX | queri, xqueri, ontolog, trec, sparql, rdf, dbpedia, wikipedia, searcher, ontology-bas |
| Inorganic Chemistry | Probability | oxid, use, reaction, surfac, activ, catalyst, complex, studi, solut, result |
| | FREX | catalyst, zeolit, voltammetri, electrocatalyt, electrocatalyst, catalyt, electrolyt, adsorpt, bimetal, cobalt |
| Intensive Care Medicine | Probability | patient, use, treatment, studi, clinic, diseas, care, infect, therapi, risk |
| | FREX | icu, dialysi, nosocomi, sepsi, hemodialysi, septic, aki, ard, ckd, pneumonia |
| Internal Medicine | Probability | patient, group, level, studi, p, associ, diabet, control, diseas, risk |
| | FREX | insulin, lipoprotein, cholesterol, leptin, triglycerid, ldl, mellitu, adiponectin, mg/dl, hdl |
| International Economics | Probability | trade, countri, foreign, effect, import, growth, intern, fdi, develop, econom |
| | FREX | renminbi, unfccc, iit, oca, exc, brics+matik, cross-bord, austria’, snga, forex |
| International Trade | Probability | trade, countri, develop, intern, export, econom, market, product, import, polici |
| | FREX | wto, fta, antidump, mne, trade, gatt, nafta, export, tariff, asean |
| Internet Privacy | Probability | inform, use, privaci, social, data, user, internet, network, person, protect |
| | FREX | calea, spam, youtub, biosimilar, medwatch, ipharmacist, p3p, banknot, rota, h2h |
| Keynesian Economics | Probability | inflat, rate, econom, unemploy, model, money, theori, use, keynesian, monetari |
| | FREX | weitzman, nairu, lsap, subjectivist, brainard, keynesian, deflat, keynes’, price-level, wage- |
| Knowledge Management | Probability | knowledg, manag, inform, use, system, develop, research, studi, technolog, paper |
| | FREX | e-govern, e-learn, tacit, organiz, telework, m-learn, e-servic, egovern, ict, coci |
| Labour Economics | Probability | employ, wage, labor, worker, effect, increas, market, use, incom, labour |
| | FREX | wage, unemploy, labour, part-tim, labor, unskil, retir, overtim, worker, pension |
| Law | Probability | law, state, right, court, legal, articl, case, one, intern, polit |
| | FREX | court, suprem, judici, lawyer, statut, litig, arbitr, attorney, tribun, liberti |
| Law and Economics | Probability | law, right, legal, properti, system, one, econom, use, state, principl |
| | FREX | lien, accessio, honesti, coas, korupc, pejovich, benhabib, rcss, bgb, chattel |
| **Library Science** | **Probability** | librari, univers, research, journal, scienc, public, publish, use, paper, inform |
| | **FREX** | librarian, librarianship, alct, archivist, ebook, librari, jstor, lists", scientometr, kirkconnel |
| **Linguistics** | **Probability** | languag, use, english, studi, word, linguist, differ, paper, mean, translat |
| | **FREX** | verb, phonolog, phonet, linguist, pronoun, sociolinguist, grammat, mandarin, morphem, adverb |
| **Literature** | **Probability** | work, one, cultur, text, literari, novel, also, critic, histori, book |
| | **FREX** | poetri, poem, poetic, poet, shakespear, literari, byron, comedi, fiction, satir |
| **Machine Learning** | **Probability** | svm, kernel, motif, p, protein, v, multi-label, nmf, seizur, wound |
| | **FREX** | twsvm, pltss, tsvm, tractogram, kenreg, besurek, svddbn, sonfn, kir3dl1, i^ |
| **Macroeconomics** | **Probability** | growth, polici, model, rate, econom, countri, effect, use, paper, inflat |
| | **FREX** | cointegr, granger, disinfl, dsge, ardl, laffer, expansionari, interest-sensit, counter-cycl, balassa-samuelson |
| **Management** | **Probability** | manag, research, develop, busi, univers, new, work, compani, educ, year |
| | **FREX** | i¾, mcvicker, hrm, jcl, evaluationen, sethi, drexel, mitch, mintzberg, cio |
| **Management Science** | **Probability** | model, research, use, develop, decis, method, approach, process, system, studi |
| | **FREX** | foresight, sisp, infrasystem, neuromarket, megaproject, mcdm, abm, geoengin, desn, mbdd |
| **Manufacturing Engineering** | **Probability** | product, manufactur, process, design, industri, technolog, system, develop, use, paper |
| | **FREX** | cim, poka–yok, mrpii, ferroalloy, foundri, servit, okp, amt, qfd, holon |
| **Marine Engineering** | **Probability** | wind, use, ship, turbin, design, system, model, oper, water, result |
| | **FREX** | hydrofoil, tow, vawt, riser, auv, rov, rudder, moor, subsea, fpso |
| **Market Economy** | **Probability** | market, econom, economi, develop, industri, enterpris, privat, capit, invest, competit |
| | **FREX** | pineappl, ivorian, cic, i919, agriculture-rel, i920, anti-dumpl, ocab, oil-for-food, interactiv |
| **Marketing** | **Probability** | market, studi, use, research, product, servic, custom, develop, manag, consum |
| | **FREX** | brand, loyalti, hotel, franchis, consumers', custom, customers', retail, tourist, b2b |
| **Mathematical Analysis** | **Probability** | equat, solut, method, problem, function, use, result, system, paper, condit |
| | **FREX** | dirichlet, cauchi, galerkin, sobolev, laplac, lipschitz, semilinear, eigenfunct, blow-up, well-posed |
| **Mathematical Economics** | **Probability** | game, model, equilibrium, theori, paper, gener, result, player, use, nash |
| | **FREX** | nash, payoff, shapley, strategy-proof, maxmin, wilki, equilibria, remarks.-, homothet, anarchi |
| **Mathematical Optimization** | **Probability** | problem, optim, algorithm, method, use, model, propos, solut, result, paper |
| | **FREX** | pso, multi-object, swarm, multiobject, subproblem, tabu, np-hard, metaheurist, nonconvex, salesman |
| **Mathematical Physics** | **Probability** | equat, theori, field, solut, gener, n, model, x, de, function |
| | **FREX** | string.-, brillouin-wign, h.j, l/r, l'impuls, duff, e2-instanton, bm.-, formalism.-, cartan' |
| **Mathematics Education** | **Probability** | student, teach, learn, use, teacher, studi, educ, school, research, mathemat |
| | **FREX** | teachers’, students’, teacher, siswa, student’, pre-servic, belajar, classroom, efl, pembelajaran |
| **Mechanical Engineering** | **Probability** | use, design, model, process, system, result, paper, method, heat, machin |
| | **FREX** | workpiec, ejector, spool, gripper, grind, compressor, gear, warpag, micropump, louver |
| **Mechanics** | **Probability** | flow, model, use, result, pressur, effect, veloc, heat, studi, simul |
| | **FREX** | reynold, unsteadi, streamwis, nusselt, flame, nozzl, swirl, laminar, inlet, vortex |
| **Media Studies** | **Probability** | cultur, new, univers, one, work, commun, year, polit, media, studi |
| | **FREX** | cheerlead, siskin, journalist, bikini, isbn, theatr, hahn, paperback, regift, shadowplay |
| **Medical Education** | **Probability** | student, educ, medic, use, studi, train, program, teach, school, research |
| | **FREX** | faculti, clerkship, interprofession, internship, postgradu, mentor, traine, osc, pbl, ipe |
| **Medical Emergency** | **Probability** | patient, hospit, medic, use, emerg, care, injuri, studi, system, health |
| | **FREX** | prehospit, ohca, cpr, resuscit, handov, ambul, triag, out-of-hospit, telemedicin, paramed |
| **Medical Physics** | **Probability** | use, radiat, clinic, dose, imag, patient, treatment, studi, develop, result |
| | **FREX** | brachytherapi, pneumon, locoregion, yb-169, intensity-modul, dosimetri, late-cours, radiologist, boost-imrt, aapm |
| **Medicinal Chemistry** | **Probability** | compound, c, co, n, h, reaction, ring, atom, die, complex |
| | **FREX** | me3si, intermolecular, pph4, intramolecular, sime, pme3, molecul, n—h···o, pph4cl, c—h···o |
| **Metallurgy** | **Probability** | alloy, temperatur, use, steel, result, process, increas, surfac, materi, coat |
| | **FREX** | alloy, austenit, martensit, corros, carbid, stainless, sinter, solder, microstructur, microhard |
| **Meteorology** | **Probability** | model, wind, use, data, result, observ, measur, studi, temperatur, forecast |
| | **FREX** | thunderstorm, typhoon, wrf, gust, meteorolog, tornado, mesoscal, trmm, radiosond, lightn |
| **Microbiology** | **Probability** | strain, isol, resist, use, bacteria, infect, studi, gene, activ, result |
| | **FREX** | aeruginosa, esbl, virul, streptococcu, faecali, imipenem, baumannii, jejuni, o157, albican |
| **Microeconomics** | **Probability** | price, model, market, paper, use, cost, product, firm, effect, result |
| | **FREX** | auction, collus, bidder, monopolist, cournot, seller, oligopoli, duopoli, oligopolist, buyer |
| **Mineralogy** | **Probability** | miner, use, sampl, studi, temperatur, result, content, water, rock, differ |
| | **FREX** | calcit, kaolinit, feldspar, pyrit, illit, hematit, aragonit, ilmenit, tourmalin, smectit |
| **Mining Engineering** | **Probability** | mine, coal, area, geolog, water, seam, method, rock, ore, use |
| | **FREX** | seam, coalfield, workfac, opencast, stope, orebodi, inrush, collieri, tabuliformi, gangu |
| **Molecular Biology** | **Probability** | cell, express, protein, gene, activ, dna, use, human, result, bind |
| | **FREX** | cdna, transfect, plasmid, blot, immunoprecipit, mrna, transactiv, luciferas, c-myc, rnase |
| **Molecular Physics** | **Probability** | structur, surfac, calcul, energi, molecul, vibrat, electron, state, defect, molecular |
| | **FREX** | v/sia, shear/mix, v-sio2, lipss, pre-arc, ssic, gnf, a-si, band-a, double-excit |
| **Monetary Economics** | **Probability** | rate, market, exchang, bank, monetari, polici, price, effect, model, paper |
| | **FREX** | monetari, ecb, fii, pass-through, mudaraba, fiat, interbank, countercycl, kwacha, trend-follow |
| **Multimedia** | **Probability** | use, learn, system, student, design, develop, paper, technolog, teach, comput |
| | **FREX** | multimedia, coursewar, moodl, e-learn, braill, playlist, conferenc, audio, actionscript, video |
| **Nanotechnology** | **Probability** | use, materi, applic, surfac, structur, nanoparticl, develop, properti, process, cell |
| | **FREX** | nanostructur, nanowir, nanotub, cnt, nanotechnolog, nanomateri, graphen, microfluid, aunp, nanoscal |
| **Natural Language Processing** | **Probability** | languag, word, text, translat, sentenc, semant, corpu, english, annot, lexic |
| | **FREX** | tagger, treebank, part-of-speech, nlp, metaschema, phrase-bas, lemmat, paraphras, stemmer, transliter |
| **Natural Resource Economics** | **Probability** | energi, develop, resourc, emiss, product, econom, industri, use, environment, increas |
| | **FREX** | non-co2, flng, biohydrogen, gudawang, bio-diesel, nsgg, ngir, rutf, non-monetari, climate-rel |
| **Neoclassical Economics** | **Probability** | theori, econom, capit, keyn, marx, product, modern, gener, economi, valu |
| | **FREX** | tiebout, radjou, sismondi, frugal, surplus-valu, nicancioglu, buchanan, anieva, sraffa, prosumpt |
| **Neuroscience** | **Probability** | neuron, activ, brain, function, studi, cell, respons, use, cortex, system |
| | **FREX** | synapt, synaps, excitatori, hippocamp, neuron, hippocampu, interneuron, postsynapt, cortex, amygdala |
| **Nuclear Chemistry** | **Probability** | use, concentr, result, acid, solut, studi, show, ph, effect, prepar |
| | **FREX** | biosorpt, biosorb, ap-al, agnp, zntcp, pseudo-second-ord, inchikey, dalapon, desfer, rofa |
| **Nuclear Engineering** | **Probability** | reactor, fuel, use, system, design, heat, oper, power, nuclear, result |
| | **FREX** | burnup, pwr, thermal-hydraul, divertor, loca, coolant, htgr, burn-up, bwr, tfe |
| **Nuclear Magnetic Resonance** | **Probability** | magnet, field, use, measur, reson, imag, result, temperatur, studi, method |
| | **FREX** | hyperfin, mossbauer, magn, quadrupol, spin-lattic, quadrupolar, spin-echo, heteronuclear, coil, gmi |
| **Nuclear Medicine** | **Probability** | use, dose, imag, patient, method, studi, measur, result, differ, treatment |
| | **FREX** | spect, fdg, vmat, pet/ct, imrt, dosimetr, ptv, isocent, sbrt, oar |
| **Nuclear Physics** | **Probability** | energi, measur, use, neutron, data, nuclear, result, reaction, experi, detector |
| | **FREX** | pion, muon, rhic, gev/c, deuteron, antiproton, mev, cern, au+au, heavy-ion |
| **Nursing** | **Probability** | care, nurs, health, patient, use, studi, servic, practic, provid, hospit |
| | **FREX** | nurs, palli, midwiv, hospic, midwiferi, carer, nurses', care, breastfeed, patient-cent |
| **Obstetrics** | **Probability** | women, pregnanc, studi, group, risk, birth, matern, use, patient, deliveri |
| | **FREX** | cesarean, trimest, caesarean, gestat, gdm, misoprostol, obstetr, pregnanc, perinat, preterm |
| **Oceanography** | **Probability** | water, sea, sediment, ocean, surfac, chang, studi, increas, concentr, area |
| | **FREX** | phytoplankton, foraminifera, upwel, benthic, gyre, zooplankton, kuroshio, plankton, foraminifer, chukchi |
| **Oncology** | **Probability** | cancer, patient, breast, surviv, treatment, studi, chemotherapi, tumor, therapi, use |
| | **FREX** | nsclc, docetaxel, neoadjuv, trastuzumab, cetuximab, chemotherapi, mcrc, non-smal, carboplatin, progression-fre |
| **Operating System** | **Probability** | system, use, applic, oper, server, file, comput, data, softwar, develop |
| | **FREX** | servlet, linux, hypervisor, unix, vmm, filesystem, xen, sharepoint, scsi, nrd |
| **Operations Management** | **Probability** | use, manag, system, cost, product, model, studi, perform, develop, process |
| | **FREX** | jit, remanufactur, qfd, subcontractor, lot-siz, tqm, kanban, retailer-l, scqi, wind-pow |
| **Operations Research** | **Probability** | model, use, system, problem, paper, method, decis, cost, develop, time |
| | **FREX** | abrf, genco, topsi, travelers', cplex, mcdm, vrp, schedule-bas, mixed-integ, macchiarini |
| **Ophthalmology** | **Probability** | eye, patient, visual, group, retin, result, studi, use, corneal, glaucoma |
| | **FREX** | iop, macular, iol, intraocular, vitrectomi, glaucoma, keratoplasti, rnfl, phacoemulsif, intravitr |
| **Optics** | **Probability** | use, optic, measur, laser, result, method, system, imag, beam, light |
| | **FREX** | grate, waveguid, interferomet, bragg, wavefront, birefring, coupler, mode-lock, femtosecond, speckl |
| **Optoelectronics** | **Probability** | use, devic, layer, film, structur, laser, optic, high, current, temperatur |
| | **FREX** | diod, gan, photodetector, heterojunct, gaa, schottki, photodiod, electroluminesc, epitaxi, heterostructur |
| **Optometry** | **Probability** | eye, visual, vision, acuiti, refract, use, patient, cataract, test, measur |
| | **FREX** | amblyopia, optometri, anisometropia, optometrist, stereopsi, logmar, spectacl, ophthalmologist, lowercas, optotyp |
| **Organic Chemistry** | **Probability** | reaction, acid, use, compound, yield, group, product, studi, activ, structur |
| | **FREX** | enantioselect, keton, aldehyd, cycliz, alken, allyl, olefin, stereoselect, enantiomer, ester |
| **Orthodontics** | **Probability** | use, patient, studi, group, measur, differ, treatment, result, method, mandibular |
| | **FREX** | malocclus, cephalometr, orthognath, orthodont, cephalogram, mandibular, tmj, overbit, incisor, osteotomi |
| **Paleontology** | **Probability** | format, speci, fossil, earli, late, new, deposit, age, basin, lower |
| | **FREX** | cambrian, conodont, biostratigraph, ordovician, brachiopod, silurian, trilobit, ammonit, miocen, devonian |
| **Parallel Computing** | **Probability** | parallel, perform, use, algorithm, comput, implement, memori, system, processor, paper |
| | **FREX** | gpu, prefetch, speedup, cuda, simd, openmp, hypercub, multiprocessor, gpgpu, multi-cor |
| **Particle Physics** | **Probability** | model, mass, decay, quark, neutrino, data, use, result, b, energi |
| | **FREX** | higg, lepton, quark, neutrino, electroweak, tev, parton, hadron, tevatron, mssm |
| **Pathology** | **Probability** | patient, cell, tumor, case, studi, diseas, cancer, use, express, clinic |
| | **FREX** | immunohistochem, lymphoma, neoplasm, immunohistochemistri, squamou, papillari, tumour, ihc, pleural, carcinoma |
| **Pattern Recognition** | **Probability** | imag, method, featur, use, propos, algorithm, base, result, classif, paper |
| | **FREX** | denois, gabor, svm, histogram, wavelet, palmprint, minutia, c-mean, k-nearest, vq |
| **Pedagogy** | **Probability** | educ, student, teacher, school, learn, studi, teach, develop, use, research |
| | **FREX** | teacher, classroom, teachers’, pedagogi, pedagog, pre-servic, preservic, literaci, curriculum, intercultur |
| **Pediatrics** | **Probability** | patient, children, studi, age, year, infant, case, diseas, group, result |
| | **FREX** | infant, wheez, varicella, pertussi, measl, breastf, breastfeed, thalassemia, vlbw, pneumococc |
| **Petroleum Engineering** | **Probability** | oil, ga, well, reservoir, product, use, pressur, water, develop, field |
| | **FREX** | wellbor, oilfield, proppant, eor, waterflood, coalb, downhol, sagd, non-darci, gas-oil |
| **Petrology** | **Probability** | reservoir, rock, ga, fault, oil, well, faci, format, data, flow |
| | **FREX** | pseudotachylit, yacheng, xu-2, qoltag, block-off, fault-block, chinl, diagenet, kupukuziman, k1q2 |
| **Pharmacology** | **Probability** | effect, drug, studi, activ, use, dose, cell, treatment, rat, result |
| | **FREX** | pharmacokinet, antinocicept, cmax, morphin, pharmacodynam, cannabinoid, anticonvuls, cyp3a4, hepatotox, concentration-tim |
| **Photochemistry** | **Probability** | reaction, complex, electron, fluoresc, use, radic, studi, state, result, abstract |
| | **FREX** | photolysi, phosphoresc, porphyrin, photophys, chromophor, photoinduc, photosystem, singlet, photochem, photoreact |
| **Physical Chemistry** | **Probability** | reaction, k, use, temperatur, calcul, energi, phase, studi, system, der |
| | **FREX** | calphad, feroxyhyt, ssz-24, eutectic-point, j·k, libh, inapnh, h2o-molekeln, mol/sup, ho/sub |
| **Physical Geography** | **Probability** | area, chang, climat, land, studi, region, temperatur, increas, use, year |
| | **FREX** | ba/cashel, ptarmigan, desertif, uraphylla, marmot, lucc, euphratica, lahar, wangkun, ghrr |
| **Physical Medicine and Rehabilitation** | **Probability** | muscl, use, studi, measur, activ, group, subject, perform, result, function |
| | **FREX** | gait, ankl, spastic, emg, quadricep, isokinet, hemipleg, hamstr, orthosi, femori |
| **Physical Therapy** | **Probability** | patient, studi, use, group, pain, result, treatment, effect, assess, p |
| | **FREX** | copd, osteoarthr, acupunctur, sf-36, migrain, percnt, rheumatolog, arthriti, musculoskelet, physiotherapi |
| **Physiology** | **Probability** | group, studi, level, age, effect, blood, differ, femal, control, physiolog |
| | **FREX** | flight.-, hfh, progin, anem, power.-, zuntz, e1c, dmr, mcpyv, ferritin |
| **Political Economy** | **Probability** | polit, state, social, polici, econom, develop, govern, parti, nation, new |
| | **FREX** | nato, democrat, democraci, authoritarian, insurg, full.click, elector, parti, superpow, soviet |
| **Polymer Chemistry** | **Probability** | polym, polymer, copolym, poli, use, group, chain, monom, temperatur, reaction |
| | **FREX** | copolymer, methacryl, copolym, polyimid, atrp, mma, polycondens, poli, polymer, styren |
| **Polymer Science** | **Probability** | polym, silk, control, structur, properti, system, materi, use, a., releas |
| | **FREX** | pysp2, trantolo, oxygen-barri, gresser, weipert, chi-652, noil, vulcaniz, making-up, aroma-barri |
| **Positive Economics** | **Probability** | econom, theori, model, social, economist, develop, discuss, argu, studi, ration |
| | **FREX** | igo, bureau-shap, transfer’, zeliz, drawing’, self-respect, frankel, heterodox, policy-mak, shackl |
| **Process Engineering** | **Probability** | process, system, use, product, energi, oper, model, design, develop, technolog |
| | **FREX** | exergi, exerget, flowsheet, polygener, hrsg, photodesmear, aspen, pervapor, gasifi, nebulis |
| **Process Management** | **Probability** | manag, process, busi, system, develop, model, enterpris, inform, servic, use |
| | **FREX** | scorecard, bpm, bpr, bpmn, scrumban, plm, uaa, omd, rpjmd, group.- |
| **Programming Language** | **Probability** | program, languag, use, system, model, paper, implement, code, specif, develop |
| | **FREX** | refactor, prolog, bytecod, haskel, debugg, acl2, cobol, rfun, prover, mizar |
| **Psychiatry** | **Probability** | patient, disord, studi, use, depress, treatment, symptom, health, mental, result |
| | **FREX** | psychiatr, schizophrenia, antipsychot, antidepress, psychot, psychosi, suicid, psychiatri, schizophren, ptsd |
| **Psychoanalysis** | **Probability** | work, one, life, psychoanalysi, way, psycholog, author, book, freud, person |
| | **FREX** | psychoanalysi, freud, jung, psychoanalyt, freudian, freud’, deadhead, fyne, jungian, fetish |
| **Psychotherapist** | **Probability** | therapi, treatment, patient, famili, therapist, psychotherapi, use, clinic, therapeut, process |
| | **FREX** | psychotherapi, therapist, hypnosi, psychotherapeut, cbt, grief, countertransfer, psychotherapist, ipt, psychodynam |
| **Public Administration** | **Probability** | govern, polici, public, polit, state, develop, administr, nation, system, educ |
| | **FREX** | senat, elector, presidenti, bureaucraci, charter, congression, parliamentari, poll, referendum, parliament |
| **Public Economics** | **Probability** | tax, polici, public, use, econom, govern, paper, effect, develop, social |
| | **FREX** | taxpay, eco-label, i.r.c, hine, nudg, schedular, hwf, sunstein, vat/gst, dehesa |
| **Public Relations** | **Probability** | research, public, social, commun, develop, manag, use, studi, educ, work |
| | **FREX** | nonprofit, csr, newsroom, fundrais, crowdfund, advocaci, non-profit, organiz, leadership, employe |
| **Pulp and Paper Industry** | **Probability** | use, process, product, oil, remov, wastewat, result, treatment, studi, effect |
| | **FREX** | kraft, anammox, bioplast, laccas, pome, bagass, sawdust, white-rot, delignif, hemicellulos |
| **Pure Mathematics** | **Probability** | space, algebra, group, gener, x, function, oper, theorem, paper, result |
| | **FREX** | c*-algebra, eisenstein, finsler, hypergroup, r-algebra, union-soft, p-set, subvarieti, near-r, artin |
| **Quantum Electrodynamics** | **Probability** | theori, field, model, gaug, use, effect, gener, function, equat, result |
| | **FREX** | yang-mil, massless, supergrav, one-loop, fermion, supersymmetr, supersymmetri, d-brane, tachyon, two-loop |
| **Quantum Mechanics** | **Probability** | quantum, state, system, use, theori, field, model, gener, function, result |
| | **FREX** | entangl, qubit, decoher, quantum, soliton, semiclass, wavefunct, wigner, bec, squeez |
| **Radiochemistry** | **Probability** | use, irradi, neutron, dose, measur, sampl, activ, determin, radiat, method |
| | **FREX** | bq, plutonium, dosimet, radiochem, bq/kg, bnct, polonium, thorium, kgi, hto |
| **Radiology** | **Probability** | patient, imag, case, use, ct, lesion, studi, diagnosi, arteri, tumor |
| | **FREX** | aneurysm, angiographi, embol, endovascular, sonographi, mediastin, contrast-enhanc, stent, vena, arterioven |
| **Real-Time Computing** | **Probability** | system, use, time, data, propos, network, perform, sensor, paper, result |
| | **FREX** | rssi, arq, dvf, macroblock, harq, stuck-at, nlo, ads-b, timeout, viewport |
| **Regional Science** | **Probability** | develop, region, citi, innov, tourism, research, paper, studi, econom, system |
| | **FREX** | rebam, placenam, rural-bas, form-funct, non-perman, laboratory’, ipalr, brussels-capit, circle-shap, kul |
| **Reliability Engineering** | **Probability** | system, reliabl, use, test, model, method, failur, power, paper, analysi |
| | **FREX** | phm, fmea, hazop, outag, mtbf, substat, switchgear, burn-in, lole, fdd |
| **Religious Studies** | **Probability** | christian, religion, religi, church, theolog, one, islam, studi, polit, god |
| | **FREX** | heschel, raju, judaism, sufism, zionism, anti-semit, catholic, rabbi, anabaptist, dharma |
| **Remote Sensing** | **Probability** | use, data, imag, measur, model, system, method, result, satellit, area |
| | **FREX** | lidar, radiomet, modi, hyperspectr, polarimetr, landsat, radianc, sar, multispectr, spaceborn |
| **Risk Analysis (Engineering)** | **Probability** | risk, system, safeti, manag, assess, develop, process, use, product, method |
| | **FREX** | haccp, qra, htr-pm, sift-proof, sva, microbicid, bepg, moniqa, grft, cipcast |
| **Seismology** | **Probability** | earthquak, seismic, fault, data, use, model, event, region, result, zone |
| | **FREX** | aftershock, earthquak, tsunami, seismolog, strike-slip, coseism, epicent, mainshock, teleseism, seismic |
| **Simulation** | **Probability** | use, simul, system, model, result, control, perform, robot, develop, design |
| | **FREX** | humanoid, exoskeleton, bipe, haptic, overtak, robot, afo, via-point, car-follow, loader |
| **Social Psychology** | **Probability** | studi, use, social, research, differ, result, effect, behavior, relationship, group |
| | **FREX** | self-esteem, stereotyp, intergroup, interperson, empathi, divorc, prejudic, accultur, self-concept, shame |
| **Social Science** | **Probability** | social, research, polit, cultur, studi, articl, develop, educ, use, paper |
| | **FREX** | sociolog, sociologist, bourdieu, haberma, chautauqua, durkheim, marxism, pequot, ecec, neoliber |
| **Socioeconomics** | **Probability** | studi, area, household, use, popul, rural, social, health, develop, urban |
| | **FREX** | kirsal, turizm, eav, kākā, haor, wpv, non-farm, jiedao, ardahan, padwcm |
| **Software Engineering** | **Probability** | softwar, system, develop, design, use, model, process, paper, applic, requir |
| | **FREX** | uml, model-driven, blender, service-ori, reusabl, soa, vph-share, idoc, cmmi, ippa |
| **Soil Science** | **Probability** | soil, water, use, content, model, differ, studi, organ, increas, result |
| | **FREX** | topsoil, macroaggreg, chernozem, loam, humu, gross-beta, humif, loami, soil, sorptiv |
| **Speech Recognition** | **Probability** | speech, use, recognit, signal, system, model, method, result, propos, perform |
| | **FREX** | phonem, asr, hmm, speech, cepstral, mfcc, formant, pronunci, triphon, vowel |
| **Statistical Physics** | **Probability** | model, system, simul, use, dynam, method, result, distribut, scale, time |
| | **FREX** | pott, scale-fre, self-avoid, langevin, ise, finite-s, ut-soi, nonextens, nonequilibrium, tricrit |
| **Statistics** | **Probability** | model, estim, use, method, data, test, distribut, sampl, statist, studi |
| | **FREX** | nonparametr, censor, semiparametr, quantil, lasso, bootstrap, jackknif, minimax, imput, two-sampl |
| **Stereochemistry** | **Probability** | structur, compound, activ, bind, complex, acid, group, two, c, r |
| | **FREX** | stereochemistri, stereoselect, nucleosid, subsit, enantiom, aglycon, stereochem, diterpen, structure–act, diterpenoid |
| **Structural Engineering** | **Probability** | model, use, structur, result, load, method, test, design, analysi, stress |
| | **FREX** | buckl, prestress, girder, bolt, stiffen, cfrp, damper, truss, crack, stiff |
| **Surgery** | **Probability** | patient, case, group, use, result, treatment, studi, surgeri, year, method |
| | **FREX** | arthroplasti, flap, femor, pedicl, postop, sutur, hematoma, arthroscop, decompress, debrid |
| **Systems Engineering** | **Probability** | system, design, develop, model, process, use, paper, product, softwar, requir |
| | **FREX** | mde, bim, avion, fieldbu, gm-vv, soss, nfr, csdp, vbe, sdec |
| **Telecommunications** | **Probability** | system, technolog, network, commun, servic, use, mobil, paper, develop, telecommun |
| | **FREX** | telecom, telecommun, intelsat, broadband, satcom, wban, fcc, umt, subscrib, tvw |
| **Theology** | **Probability** | theolog, god, christian, church, articl, one, work, also, new, studi |
| | **FREX** | radd, eucharist, trinitarian, ecclesiolog, sermon, maimonid, christolog, mennonit, qur'an, colonna |
| **Theoretical Computer Science** | **Probability** | use, model, algorithm, system, comput, problem, paper, graph, network, propos |
| | **FREX** | hash, cryptanalysi, lineariz, cryptosystem, bdd, cipher, automata, plaintext, zero-knowledg, diffie-hellman |
| **Theoretical Physics** | **Probability** | theori, physic, quantum, model, univers, gener, one, use, discuss, time |
| | **FREX** | mcat, einstein', antiscalar, lqc, pii, gsl, delayed-choic, jarzynski, neurcitosti, w3u |
| **Thermodynamics** | **Probability** | heat, temperatur, model, use, result, experiment, system, transfer, flow, pressur |
| | **FREX** | nanofluid, nusselt, boil, prandtl, subcool, supercool, undercool, thermophys, vapor-liquid, superh |
| **Topology** | **Probability** | space, group, x, n, gener, topolog, result, algebra, show, g |
| | **FREX** | submanifold, hypersurfac, homeomorph, cohomolog, riemannian, homotopi, holomorph, codimens, quiver, indecompos |
| **Toxicology** | **Probability** | exposur, use, effect, toxic, studi, test, concentr, control, level, result |
| | **FREX** | deltamethrin, insecticid, pyrethroid, cypermethrin, diazinon, genotox, chlorpyrifo, imidacloprid, neem, ddvp |
| **Traditional Medicine** | **Probability** | extract, use, medicin, activ, effect, studi, plant, group, tradit, method |
| | **FREX** | herbal, rhizoma, herb, ethnopharmacolog, decoct, ekstrak, phytochem, ethnobotan, chm, ayurved |
| **Transport Engineering** | **Probability** | traffic, transport, use, system, road, model, vehicl, develop, studi, paper |
| | **FREX** | pedestrian, freeway, lane, freight, highway, passeng, roundabout, ridership, rail, toll |
| **Urology** | **Probability** | patient, group, prostat, renal, bladder, urinari, studi, treatment, use, p |
| | **FREX** | prostatectomi, transurethr, bph, ureter, detrusor, gfr, urodynam, bladder, intraves, turp |
| **Veterinary Medicine** | **Probability** | infect, studi, sampl, anim, use, diseas, group, differ, preval, cattl |
| | **FREX** | helminth, brucellosi, seropreval, teat, zoonot, eimeria, tick, wnv, contortu, strongyl |
| **Virology** | **Probability** | infect, viru, vaccin, use, cell, antibodi, studi, viral, detect, patient |
| | **FREX** | viru, hiv-1, virus, viral, virion, hbv, hcv, capsid, env, hsv-1 |
| **Visual Arts** | **Probability** | art, work, music, artist, use, design, new, cultur, paint, imag |
| | **FREX** | sculptur, veld, lipstick, costum, b-boy, conlon, salon, brocad, artist, dega |
| **Waste Management** | **Probability** | use, wast, process, product, system, energi, result, fuel, water, studi |
| | **FREX** | boiler, bioga, sludg, inciner, flue, msw, compost, wast, gasif, desulfur |
| **Water Resource Management** | **Probability** | water, resourc, irrig, use, river, area, system, develop, manag, suppli |
| | **FREX** | meus, micro-irrig, km~2, mkayel, lulcc, water-sav, dongp, hemavathi, geum-riv, river-basin |
| **Welfare Economics** | **Probability** | de, la, le, en, e, que, da, et, lo, el |
| | **FREX** | monetaria, empresa, majoritarian, contabilidad, gazdasagi, oepnv, fenntarthato, cout, inflacao, gobierno |
| **World Wide Web** | **Probability** | web, servic, inform, use, user, system, data, develop, paper, applic |
| | **FREX** | web, metadata, browser, html, orcid, hypertext, ajax, w3c, e-book, uddi |
| **Zoology** | **Probability** | speci, new, genu, describ, sp, two, nov., morpholog, n., group |
| | **FREX** | nov., n.sp., subgenu, blakea, penney, redescrib, seta, synonymi, almeda, sp.n |

**Supplementary Table S2.** Ordinary Least Squares (OLS) regression examining the impact of paper and grant interdisciplinarity on paper success, measured by log-transformed 10-year citation counts (C10). Fixed effects for publication year and field are included as dummy variables.

| | Model (1) | Model (2) | Model (3) |
|---|---|---|---|
| Paper Interdisciplinarity (Reference) | 0.063*** (0.002) | | 0.060*** (0.002) |
| Avg. Grant Interdisciplinarity | -0.054*** (0.002) | | -0.052*** (0.002) |
| Avg. Grant-Grant Distance | -0.004* (0.002) | | -0.027*** (0.002) |
| Number of Authors | | 0.160*** (0.002) | 0.160*** (0.002) |
| Number of Grants | | 0.066*** (0.002) | 0.064*** (0.002) |
| Number of Institutes | | -0.006*** (0.002) | -0.003 (0.002) |
| Number of Funding Countries | | 0.011*** (0.002) | 0.013*** (0.001) |
| Total Funding Amounts (USD) | | 0.140*** (0.002) | 0.145*** (0.002) |
| Dummy - Year | Yes | Yes | Yes |
| Dummy - Discipline | Yes | Yes | Yes |
| Constant | 2.039*** (0.115) | 2.367*** (0.113) | 2.447*** (0.113) |
| N | 515796 | 515796 | 515796 |
| $R^2$ | 0.097 | 0.135 | 0.139 |

Standard errors in parentheses.
* p<.05, ** p<.01, *** p<.001

**Supplementary Table S3.** Negative Binomial regression examining the impact of paper and grant interdisciplinarity on paper success, measured by 10-year citation counts (C10). Fixed effects for publication year and field are included as dummy variables.

| | Model (1) | Model (2) | Model (3) |
|---|---|---|---|
| Paper Interdisciplinarity (Reference) | 0.052*** (0.002) | | 0.052*** (0.002) |
| Avg. Grant Interdisciplinarity | -0.052*** (0.002) | | -0.053*** (0.002) |
| Avg. Grant-Grant Distance | -0.003 (0.002) | | -0.029*** (0.002) |
| Number of Authors | | 0.137*** (0.002) | 0.140*** (0.002) |
| Number of Grants | | 0.069*** (0.002) | 0.067*** (0.002) |
| Number of Institutes | | 0.008*** (0.002) | 0.011*** (0.002) |
| Number of Funding Countries | | 0.018*** (0.001) | 0.020*** (0.001) |
| Total Funding Amounts (USD) | | 0.152*** (0.002) | 0.154*** (0.002) |
| Dummy - Year | Yes | Yes | Yes |
| Dummy - Discipline | Yes | Yes | Yes |
| Constant | 3.246*** (0.032) | 3.466*** (0.032) | 3.467*** (0.032) |
| N | 515796 | 515796 | 515796 |
| Pseudo $R^2$ | 0.053 | 0.093 | 0.097 |

Standard errors in parentheses.
* p<.05, ** p<.01, *** p<.001